\documentclass[aps,amssymb,amsmath,amsfonts,twocolumn]{revtex4}
\usepackage{graphicx}
\usepackage{bm,bbm}
\usepackage{color}
\usepackage{verbatim} 
\usepackage{pifont}
\def\club{\ding{168}}
\def\diamond{\color{red}\ding{169}}
\def\wdiamond{\color{white}\ding{169}}

\def\spade{\ding{171}}

\graphicspath{{graphics/}}

\begin{document}
\title{Quantum version of the Euler's problem:  a geometric perspective}

\medskip

\author{Karol {\.Z}yczkowski$^{1,2}$}

\affiliation{${}^1$Institute of Theoretical Physics,
    Jagiellonian University, ul. {\L}ojasiewicza 11, 30-348 Krak{\'o}w, Poland}

\affiliation{${}^2$Center for Theoretical Physics, Polish Academy of Sciences, Aleja Lotnik{\'o}w 32/46, 02-668 Warsaw, Poland}

\medskip

\date{December 23, 2022}

\begin{abstract}
The classical combinatorial problem of $36$ officers has no solution,
as there are no Graeco-Latin squares of order six.
The situation changes if one works in a quantum setup and allows for
superpositions of classical objects and admits entangled states.
We analyze the recently found solution to the quantum version of the Euler's 
problem from a geometric point of view. The notion of 
a {\sl non-displaceable manifold} embedded in a larger space is recalled.
This property implies that any {\sl two} copies of such a manifold, 
like two great circles on a sphere, do intersect. 
Existence of a quantum Graeco-Latin square of size six,
equivalent to a maximally entangled state of four subsystems
with $d=6$ levels each, implies that {\sl three} copies of the
manifold $U(36)/U(1)$ of maximally entangled states of the $36\times 36$ system,
embedded in the complex projective space ${\mathbbm C}P^{36\times 36 -1}$,
do intersect simultaneously at a certain point. 
\end{abstract}

\maketitle

{\sl Dedicated to the memory of Bogdan Mielnik} 

\section{Introduction}

Quantum information processing takes place in laboratory \cite{Peres}. 
To describe such a process from a theoretical point of view 
it is convenient to use the notion of a density matrix $\rho$
and the set $\Omega_d$ of all quantum states of size $d$. 
As emphasized in pioneering papers by Mielnik \cite{Mie68,Mie81},
the most important feature of this set is its convexity.

Extremal points of the convex set $\Omega_d$
are formed by projectors on the pure quantum states,
$P_{\psi}=|\psi\rangle \langle \psi|$.
Any  pure state corresponds to a 
vector in a complex Hilbert space,  $|\psi\rangle \in {\cal H}_d$,
normalized as $ \langle \psi|\psi\rangle =1$.
As the global phase  $\alpha$ is not measurable
it is sufficient to analyze equivalence classes,
$|\psi\rangle \sim e^{i \alpha} |\psi\rangle$,
which form a complex projective space
${\mathbbm C}P^{d-1}$ of $2(d-1)$ real dimension.

In the simplest case of a single-qubit system,
 $d=2$, the space of pure states forms the Bloch sphere,
 $S^2={\mathbbm C}P^{1}$. Although for higher $d$
 the complex projective space ${\mathbbm C}P^{d-1}$ 
 is not equivalent to a hypersphere, it enjoys
 the same homogeneity property: 
There are no priviliged points as all its points are alike. 

The situation changes if  a certain structure is imposed to the system.
Consider, for instance, a nine dimensional system described by a state 
$|\psi\rangle \in {\cal H}_9$. Assume that it is physically justified
to distinguish two subsystems of size three,
and identify a  two-qutrit system ${\cal H}_9= {\cal H}_3 \otimes  {\cal H}_3
=  {\cal H}_A \otimes  {\cal H}_B$, where integers in indices
 denote the dimensionality of subsystems, which are labeled by capital letters.
With respect to a given splitting of the composed Hilbert space,
 ${\cal H}_{AB}= {\cal H}_A \otimes  {\cal H}_B$,
one defines {\sl separable} states, which have the tensor product structure,
$|\psi_{\rm sep}\rangle =|\phi_A\rangle \otimes |\phi_B\rangle$.
Note that a partial trace of the projector 
$|\psi_{\rm sep}\rangle \langle \psi_{\rm sep} |$ is pure, so its rank $r=1$.

All other states, with reduced state $\sigma_A={\rm Tr}_B 
|\psi_{AB}\rangle \langle \psi_{AB}|$
 of rank $r\ge 2$  are called {\sl entangled}.
If the partial trace $\sigma_A$ is proportional to identity,
the state $|\psi_{AB}\rangle$ is called {\sl maximally entangled}.
These states, for a two-qubit system called {\sl Bell states}, play a crucial
role in the theory of quantum information processing \cite{Nielsen}.
The structure of the manifolds of separable and maximally entangled states pure
for a bi-partite $d \times d$ systems is well understood \cite{BZ17}.

On the other hand, for multipartite systems, described 
in the Hilbert space with several factors,  
 ${\cal H}_A \otimes  {\cal H}_B \otimes \cdots \otimes {\cal H}_{N}$,
the  notion of extremal entanglement is not unique \cite{GBP98,MW02}, 
as it depends on the measure selected  \cite{Sc04,FFPP08}.  

If the number $N$ of subsystems is even, one defines
{\sl absolutely maximally entangled}  (AME) states, 
as these which are maximally entangled for 
all symmetric partitions \cite{HCLRL12}.
In the case of $N=4$ subsystems $ABCD$
one needs to verify that three bipartite reduced states,
$\sigma_{AB}$, $\sigma_{AC}$ and  $\sigma_{AD}$ 
are maximally mixed.
In a seminal paper of Higuchi and Sudbery
it was proved that in the case of a four-qubit system, $N=4$ and $d=2$,
there are no AME states \cite{HS00}. However, for some other 
values of the parameters $N$ and $d$, such pure quantum states
with extremal properties are known \cite{Sc04,FFPP08}
and they find applications in certain quantum protocols \cite{HCLRL12}.
  
There are useful relations between 
classical combinatorial designs \cite{CD07} and 
strongly entangled multipartite quantum states \cite{GZ14,GALR15}.
The existence of two mutually orthogonal Latin squares (MOLS) \cite{JCD01}
 of order $d$, also called {\sl Graeco-Latin squares},
implies existence of AME states of $N=4$ subsystems with $d$ levels each,
written AME$(4,d)$.
As these designs exist for $d=3,4,5$ and any $d\ge 7$ -- see \cite{BSP60,JCD01},  
it is possible to construct states AME$(4,d)$ for these dimensions.

In dimension $d=6$, the smallest number which is neither a prime nor a power of prime, there are no Graeco-Latin squares (another name of MOLS),
as conjectured by Euler and proved by Tarry  \cite{Tarry,Stin84}.
Hence the problem, whether there exist AME$(4,6)$ state
was open  until very recently \cite{HW_table,Ri20,HRZ22}. 

The goal of the present contribution is to 
analyze the recent solution of the quantum analogue
of the Euler problem of $36$ officers \cite{RBBRLZ22,ZBRBRL22}
from a geometric  perspective. In the next section 
the structure of the manifolds of separable and maximally 
entangled states of a bipartite system is reviewed.
The notion of non-displaceable manifolds is recalled
in Sec.  \ref{sec_nondisp} and a conjecture 
concerning mutually maximally entangled states
for a $d \times d$ system is presented.
Absolutely maximally entangled states for multipartite systems are  
discussed in Sec. \ref{sec_AME},
 their link to orthogonal Latin squares is
analyzed in Sec. \ref{sec_classOLS},
while quantum orthogonal Latin squares 
are recalled in Sec. \ref{sec_quantOLS}.

In Sec. \ref{sec_deq6} we show some consequences
of the fact that a solution of the quantum version of the problem
of $36$ officers of Euler does exist.
It implies existence of an AME state of a four-partite system $ABCD$
with $d=6$ levels each, which in turn implies
 that the intersection of  three copies of manifolds $PU(36)=U(36)/U(1)$ of maximally entangled bi-partite states,
corresponding to partitions $AB|CD$ and $AC|BD$ and $AD|BC$,
embedded in the complex projective space ${\mathbbm C}P^{36\times 36 -1}$,
is not empty. As such four-qubit AME states do not exist,
 for $d=2$ analogous three manifolds $U(4)/U(1)$ embedded in ${\mathbbm C}P^{4 \times 4-1}$ 
do not intersect in a single point.
Key mathematical notions used in this work are listed in Appendix 
\ref{sec_glos}.

\section{Separable and maximally entangled states of a bipartite
$d \times d$ quantum system}
\label{sec_separable}

Consider a pure quantum state
of a $d \times d$ bipartite system, $|\psi\rangle \in {\cal H}_A \otimes {\cal H}_B$.
Let ${\cal P}_{d^2} = {\mathbbm C}P^{d^2-1}$
denote the  space of pure quantum states of this system.
It has $2(d^2-1)$ real dimensions.
Introducing a product basis $|i,j\rangle = |i\rangle \otimes |j\rangle$
with $i,j=1,\dots d$, one can write any state as a double sum,
arranging the coefficients into a complex matrix $C$,
\begin{equation}
|\psi\rangle = \sum_{i,j=1}^d C_{ij}|i\rangle \otimes |j\rangle
= \sum_{i=1}^r \sqrt{\lambda_i} |i'\rangle \otimes |i''\rangle,
\label{state2}
\end{equation}
where $r\le d$ denotes the rank of $C$.
The second form with a single sum, known as the {\sl Schmidt decomposition} 
of $|\psi\rangle$, involves singular values $\sqrt{\lambda_i} $ of the matrix $C$.
Thus  $\lambda_i \ge 0$ represent eigenvalues of a positive semidefinite 
 matrix $CC^{\dagger}$, while the rotated basis $|i',i''\rangle$,
is determined by the eigenvectors of $CC^{\dagger}$ and $C^{\dagger}C$.
 The standard  normalization condition, $||\psi||^2=1$,
  implies  $||C||^2={\rm Tr}CC^{\dagger}=1$,
 so that the Schmidt coefficients 
 form a probability vector $\Lambda$,
 with  $\sum_{i=1}^d  \lambda_i=1$.
 
 A separable state  $|\psi_{\rm sep}\rangle$,
 is by definition represented by the ordered vector
 $\Lambda=(\lambda_1 \ge \lambda_2 \ge \cdots, \ge \lambda_d)=(1,0,\dots,0)$. 
 Any state with $r\ge 2$ positive Schmidt
 coefficients is called {\sl entangled},
 so the scaled rank of a matrix, $(r(C)-1)/(d-1)$, can serve
  as a rough measure of entanglement. To get a better characterization
 one  often uses the {\sl entanglement entropy}, $S(|\psi\rangle)=S(\Lambda)
 = - \sum_{i=1}^d  \lambda_i \ln \lambda_i$,
that vanishes for separable states, $S(|\psi_{\rm sep}\rangle)=0$.
Any bipartite state  with $\Lambda=(1,1,\dots,1)/d$ is called
 {\sl maximally entangled},
 as it corresponds to the maximal entanglement entropy, 
 $S(|\psi_{\rm max}\rangle)=\ln d$.
 In this case the partial trace of the projector, $\rho_{\psi}=|\psi\rangle \langle \psi|$,
 is maximally mixed,
 ${\rm Tr}_B \rho_{\psi} = CC^{\dagger} = {\mathbbm I}/d$,
 and the same is true for the dual reduction,
 ${\rm Tr}_A \rho_{\psi} = C^{\dagger}C = {\mathbbm I}/d$.
 Hence an equivalent condition for maximal degree of mixing is that
 the matrix of coefficients is proportional to a unitary, $C=V/\sqrt{d}$ such that
 $VV^{\dagger}= {\mathbbm I}_d$. 

 Let $U_{\rm loc}=U_A\otimes U_B$ denote a {\sl local unitary} (LU),
 as $U_A, U_B \in U(d)$ and  $U_{\rm loc} \in U(d^2)$.
The Schmidt decomposition  (\ref{state2}) implies that the 
Schmidt vector $\Lambda$ and the entropy function $S(\Lambda)$ 
are invariant with respect to local unitaries:
two locally equivalent bi-partite states, 
$|\psi_{AB}\rangle \sim_{LU}
|\psi'_{AB}\rangle=U_A\otimes U_B |\psi_{AB}\rangle$,
 share the same amount of entanglement.
  
 In the simplest  case of a two-qubit system, $d=N=2$,
the maximal entanglement  is characteristic to the Bell state,
\begin{equation}
|\psi_{\rm Bell}\rangle = \frac{1}{\sqrt{2}} \bigl( |1,1\rangle + |2,2\rangle \bigr),
\label{Bell_state}
\end{equation}  
 which exhibits perfect correlations between results of measurement in subsystems
 $A$ and $B$. It is easy to check that both partial traces are maximally mixed,
 so having perfect knowledge about the state of a composed system one has no information about its parts. Observe that the state
 $|\psi_{\rm H}\rangle = ( |1,1\rangle + |1,2\rangle +|2,1\rangle - |2,2\rangle)/2$
 is maximally entangled, as the matrix $C$ is proportional to the unitary 
 Hadamard matrix $H_2$,
 so it is locally equivalent to the Bell state. 
  
The set of separable states,  written ${\cal P}_{\rm sep}$,
contains states of the form,
$(U_A\otimes U_B) ( |\phi_A\rangle\otimes |\phi_B\rangle)$,
which correspond to independent events in the classical theory of probability.
The set of separable states of real dimension $4(d-1)$
is thus given by the Cartesian product, which forms the Segre embedding,
\begin{equation}
{\cal P}_{\rm sep}=
 {\mathbbm C}P^{d-1} \times  {\mathbbm C}P^{d-1}
 \subset  {\mathbbm C}P^{d^2-1}.
\label{separ}
\end{equation}
In the case of a two-qubit system, $d=2$,
one has $ {\mathbbm C}P^{1} = S^2$,
so it is just an embedding of the product of two spheres,
$S^2 \times S^2  \subset  {\mathbbm C}P^{3}.$

Any maximally entangled state can be obtained 
from the generalized Bell state by a local rotation,
$|\psi_{\rm max}\rangle=
(U_A \otimes {\mathbbm I})\frac{1}{\sqrt{d}}\sum_{i=1}^d |i,i\rangle $.
An equivalent expression is obtained by applying the local
transformation  $({\mathbbm I}\otimes U_B)$.
As any global phase, $e^{i \alpha} \in U(1)$,
does not influence the state,
 the set of maximally entangled states, 
  of real dimension $d^2-1$, has the structure 
  of the manifold corresponding to the projective unitary group   \cite{BZ17},
\begin{equation}
{\cal P}_{\rm max} \cong PU(d)  = U(d) /U(1) 
 \subset  {\mathbbm C}P^{d^2-1}.
\label{entan}
\end{equation}

\begin{figure}[htbp]
\begin{center}
\includegraphics[width=0.51\textwidth]{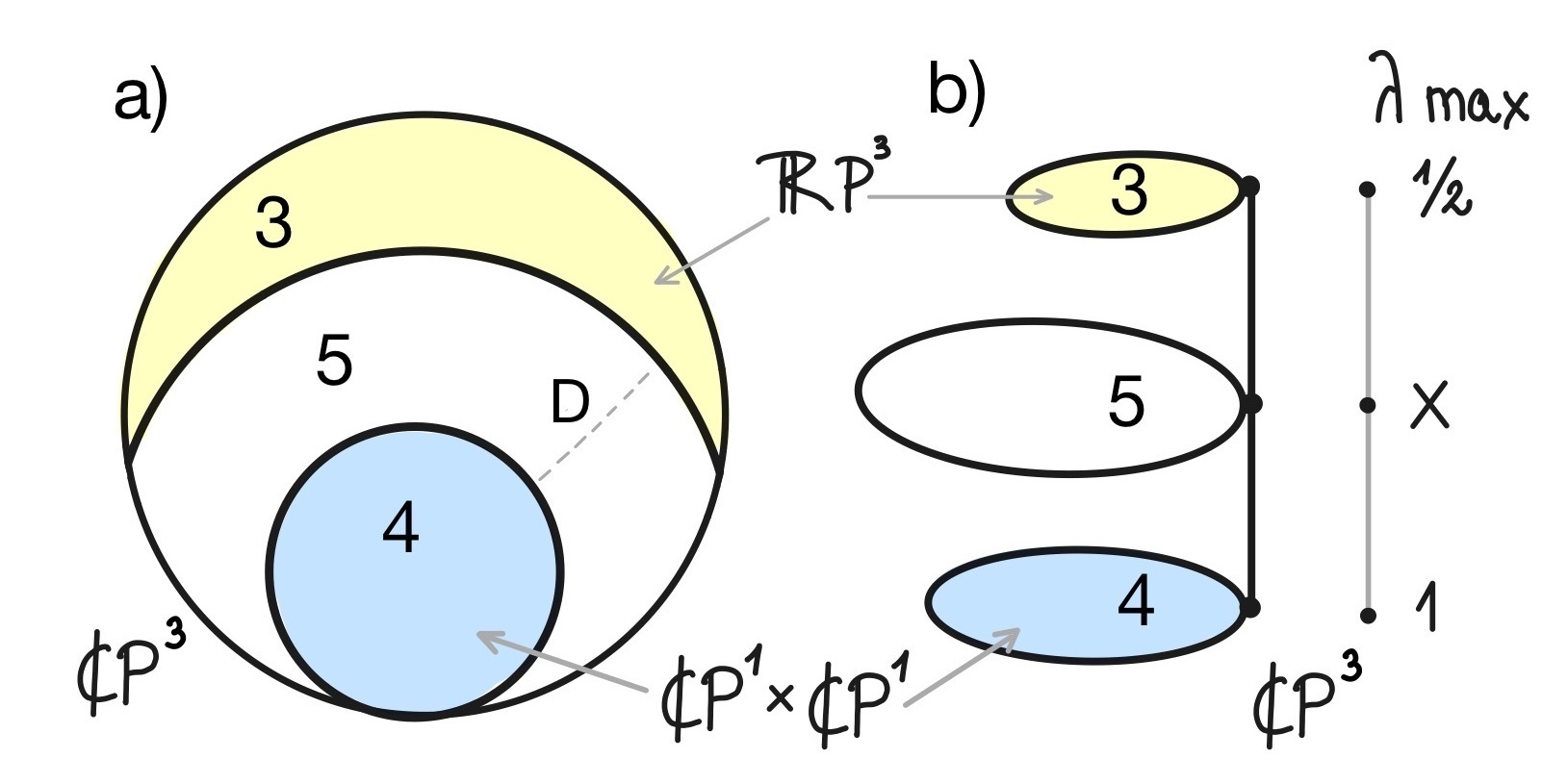} 
\caption{Set of pure states for two-qubit system,
${\cal P}_{4} = {\mathbbm C}P^{3}$.
a) Sketch shows the subset of maximally entangled states,
${\cal P}_{\rm max}  = PU(2)= U(2) /U(1) = {\mathbbm R}P^{3}$,
 located at the maximal Fubini-Study distance
  $D= \arccos(1/\sqrt{2})=\pi/4$ from the
 product of two spheres, forming the set of separable states,
${\cal P}_{\rm sep}  = S^2 \times S^2 \subset {\mathbbm C}P^{3}$.
Panel b) shows its foliation into local orbits labeled by  
the Schmidt vector $\Lambda=(x,1-x)$.
For a generic value of $x$ the local orbit
 has 5 real dimensions.
}
\label{fig:2qubit}
\end{center}
\end{figure}

 Figure \ref{fig:2qubit}
 shows a scheme of the $6$-dimensional space 
 ${\cal P}_{4}$  of pure  states for a two-qubit system.
 A simple two dimensional sketch shown in  Fig. \ref{fig:2qubit}a cannot be precise:
 on one hand the set  ${\cal P}_{\rm max}$
 of maximally entangled states is located `as far' as possible
 from the Cartesian product of two Bloch spheres,
 forming the set ${\cal P}_{\rm sep}$ of separable states.
 On the other hand, for any dimension $d$,
 the set ${\cal P}_{\rm max}$  should not
 be situated `at the boundary' of the set ${\cal P}_{d^2}$,
 as it corresponds to the center  $\Lambda_*=(1,1,\dots,1)/d$ 
 of the simplex of the Schmidt vectors, while separable states
 form an orbit which connects its all corners,
 $\Lambda=(1,0,\dots,0)$. 
 
 Fig. \ref{fig:2qubit}b
 shows a foliation of ${\mathbbm C}P^{3}$
 generated by the Schmidt coefficients:
 any point $x$ from $[1/2,1]$ generates
 an orbit of locally equivalent states,
 \begin{equation}
 |\psi_x\rangle = (U_A\otimes U_B) [ \sqrt{x}|11\rangle +  \sqrt{1-x}|22\rangle].
\label{orb2}
\end{equation}
A  generic orbit corresponding to $x\in (1/2,1)$ has 
$5$ dimensions and the structure
 $\frac{U(2)}{[U(1)]^2} \times {\mathbbm R}P^{3}$.
 For $x=1/2$ the manifold of maximally entangled states,
${\cal P}_{\rm max} = U(2) /U(1)$, has $3$ dimensions,
while the stratum of separable states 
${\cal P}_{\rm sep}  = {\mathbbm C}P^{1} \times {\mathbbm C}P^{1} $
 at $x=1$ has $4$ dimensions.

For any higher dimension $d$ it is not simple to understand 
the geometry
of the space ${\cal P}_{d^2} = {\mathbbm C}P^{d^2-1}$
 of pure states of a two-qudit system,
 with embedded subspaces
 ${\cal P}_{\rm sep}= {\mathbbm C}P^{d-1} \times {\mathbbm C}P^{d-1}$
  of separable states and 
  ${\cal P}_{\rm max}= PU(d)=U(d)/U(1)$
  of maximally entangled states.
  One can apply the notion of the numerical range of an operator of order $d^2$,
    which gives a projection of the 
    entire projective space ${\mathbbm C}P^{d^2-1}$
      onto a plane \cite{DGHMPZ11}
    and their variants including  entangled and separable numerical ranges \cite{PMGDHZ12},
    but these projections do depend on the operator selected
    and do not always lead to a clear visualization of the manifold projected.
    
\begin{figure}[htbp]
\begin{center}
\includegraphics[width=0.47\textwidth]{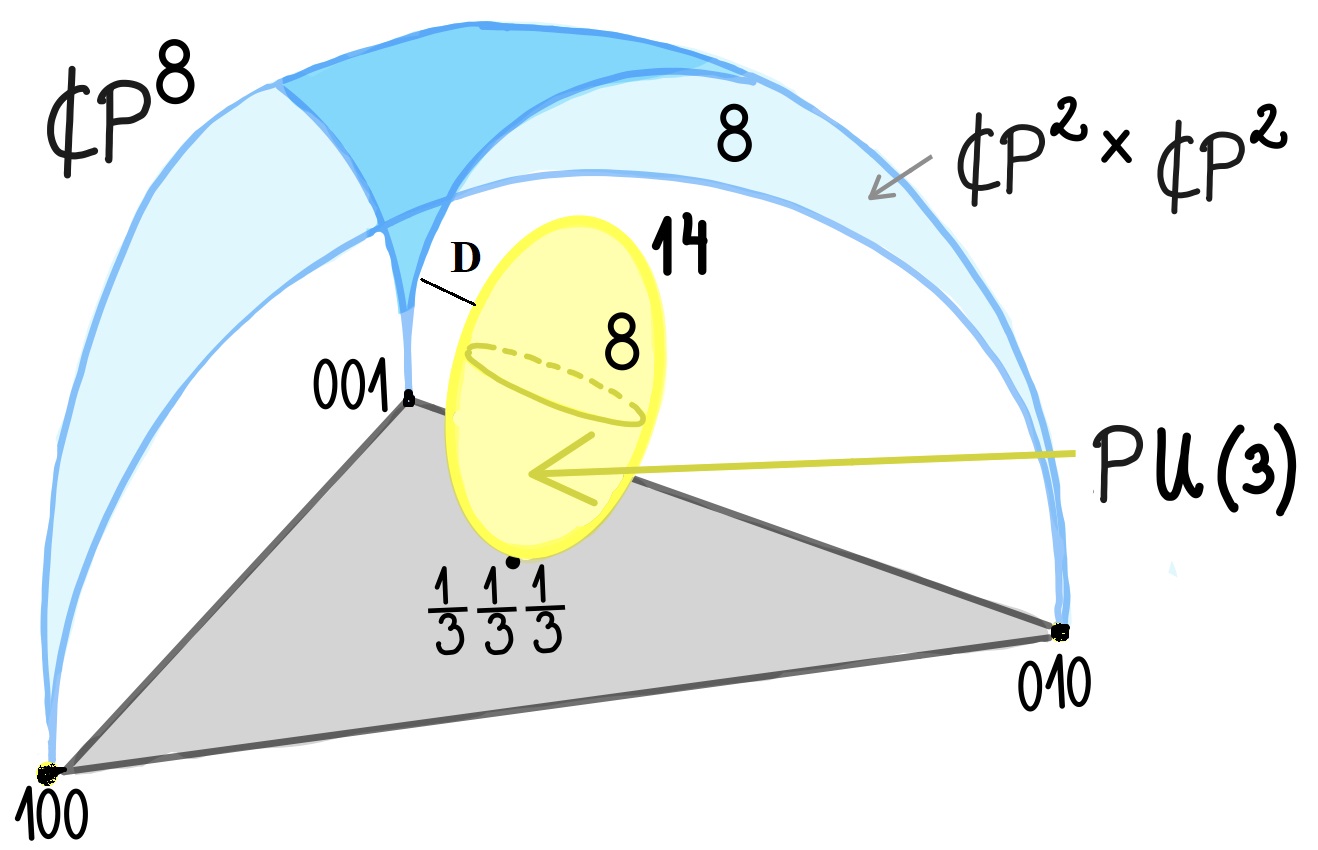}
\caption{Foliation of  $16$ dimensional set ${\cal P}_{9} = {\mathbbm C}P^{8}$
of pure states for a $3 \times 3$ system induced by the 
simplex $\Delta$ of Schmidt vectors: the set of product states,
${\mathbbm C}P^{2} \times {\mathbbm C}P^{2}$,
connecting three corners of the simplex,
has $8$ dimensions.  The antipodal subset, 
${\cal P}_{\rm max}  = PU(3)=U(3)/U(1)$, of maximally entangled states,
which emerges from the center of the simplex is of the same dimensionality.
Any generic orbit  is $14$ dimensional. 
}
\label{fig:2qutrit}
\end{center}
\end{figure}

    Therefore,   inspired by the recent work \cite{EHMS21}, which presents a 
    nice way to elucidate the structure of the eight  dimensional set $\Omega_3$ of mixed states of a qutrit, in Fig. \ref{fig:2qutrit}
   we set $d=3$ and  show a sketch of the foliation of the space 
     ${\cal P}_{9} = {\mathbbm C}P^{8}$ of pure states for two qutrits
      into local orbits.
    Consider the simplex of Schmidt coefficients of a bipartite state 
    $|\psi_{AB}\rangle \in {\cal H}_9$,
     which  describes also the spectrum of both partial traces.
     The orbit of separable states,
     ${\cal P}_{\rm sep}= {\mathbbm C}P^{2} \times {\mathbbm C}P^{2}$,
      connecting all three corners of the simplex, has $4+4=8$ real dimensions.
      In this case also the manifold
         ${\cal P}_{\rm max}= PU(3)$ of maximally entangled states,
           originating from the center of the simplex $\Lambda_*=\frac{1}{3}(1,1,1)$,
        has $9-1=8$ real dimensions. 
        Analyzing foliation of the projective space ${\cal P}_{9}$ 
       with respect to the base formed by the simplex $\Delta_3$ of the Schmidt coefficients   
        the manifold  ${\cal P}_{\rm max}$ sits `just its the center' 
        and its  distance to  ${\cal P}_{\rm sep}$ reads  $D= \arccos(1/\sqrt{3})$
         -- see Fig. \ref{fig:2qutrit}.        
          A generic orbit, stemming from a typical point of the simplex 
            with rank two, $(x,y,1-x-y)$,
            contains states
          \begin{equation}
 |\psi_{x,y}\rangle = (U_A\otimes U_B) [ \sqrt{x}|11\rangle + 
  \sqrt{y}|22\rangle] +
  \sqrt{1-x-y}|33\rangle].
\label{orb3}
\end{equation}
          Such an orbit of locally equivalent states   
            has the structure $[U(3)/U(1)] \times [U(3)/U(1)^{\times 3}]$
              and  $8+6=14$ dimensions \cite{BZ17}.

\section{Non-displaceable manifolds and mutually entangled states}
\label{sec_nondisp}

A pure state of a system described in a Hilbert space
of dimension $d^2$
is called separable with respect to a particular splitting
of the Hilbert space,
 ${\cal H}_{d^2}=  {\cal H}_d \otimes {\cal H}_d
 = {\cal H}_A \otimes {\cal H}_B$,
 if it has a product form, $|\phi_A\rangle \otimes |\phi_B\rangle$.
 A state $|\psi_{\rm sep}\rangle$, separable with 
 respect to the partition ${\cal H}_A \otimes {\cal H}_B$,
  needs not to be separable with respect to any other partition,
 ${\cal H}_{d^2}=  {\cal H}_{A'} \otimes {\cal H}_{B'}$,
 obtained from the initial one by a global unitary transformation
 $U\in U(d^2)$. The same property concerns maximal entanglement:
  a state $|\psi_{\rm max}\rangle$, maximally entangled
 with respect to the partition $A|B$ is not necessarily 
  maximally entangled with respect to another partition $A'|B'$.
  
  We are going to analyze {\sl mutually entangled} states \cite{PRCPZ15},
  which are maximally entangled with respect to two given partitions
   $A|B$ and $A'|B'$.
   For this purpose, it is convenient to use the notion of 
   {\sl non-displaceable manifolds}. 
   Such a manifold embedded in a larger space
    has a particular feature that it cannot be displaced into any other position,
in a way, that the original manifold and the displaced one 
do not intersect \cite{Cho04,Tam08}.

\begin{figure}[htbp]
\begin{center}
\includegraphics[width=0.22\textwidth]{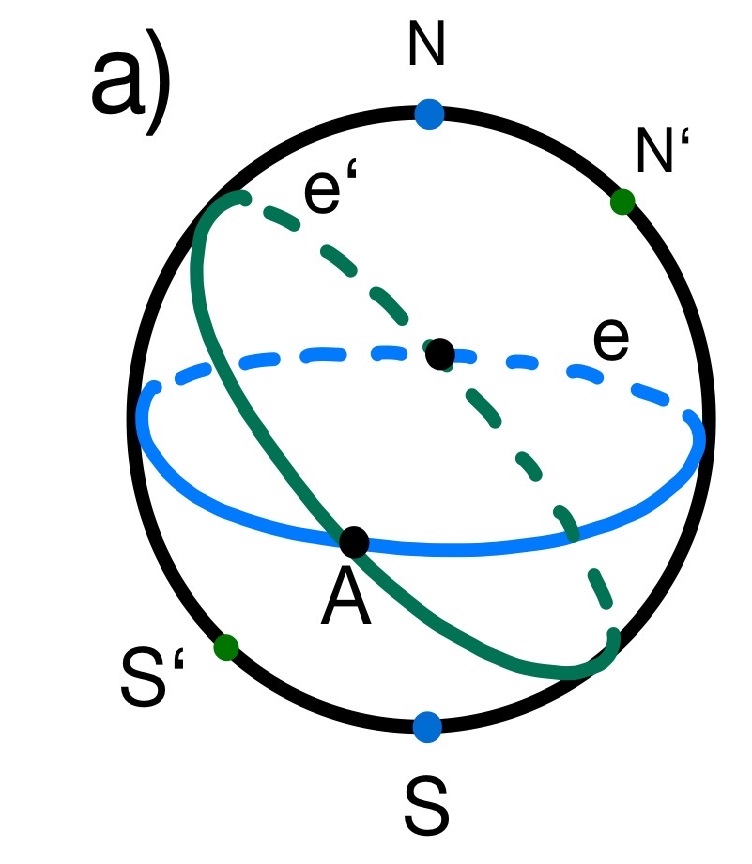} \
 \includegraphics[width=0.24\textwidth]{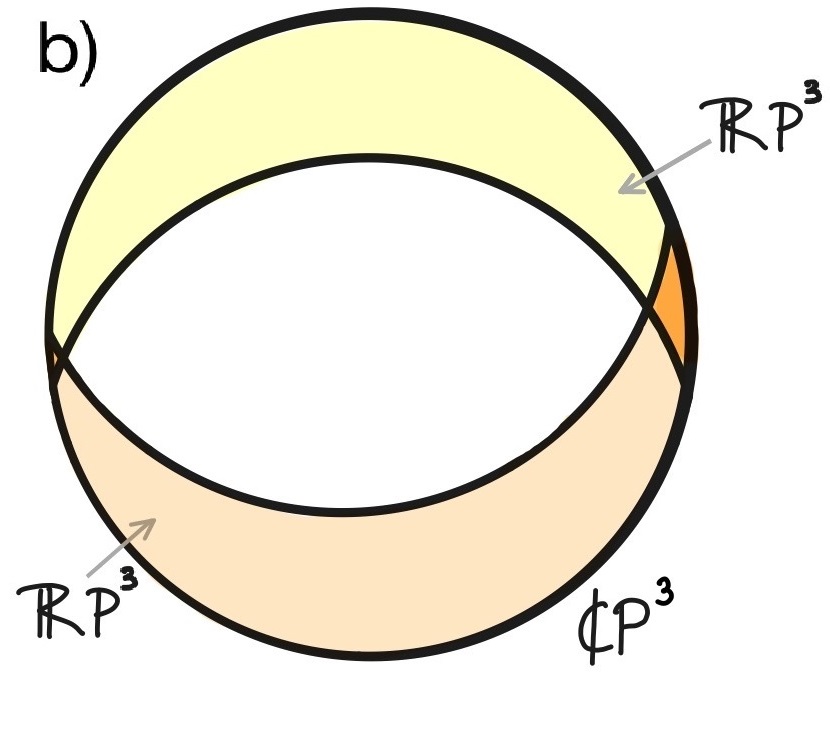}
\caption{a) Equator $S^1$ is non-displaceable in $S^2$
as two great circles on a sphere  $e$ and $e'$ do intersect;
b) the set of maximally entangled two-qubit states,
${\cal P}_{\rm max}  = PU(2)= 
{\mathbbm R}P^{3}$,
is non-displaceable in the
set of all pure states, ${\cal P}_{4} = {\mathbbm C}P^{3}$,
so mutually entangled states do exist in this case.}
\label{fig:3nondis}
\end{center}
\end{figure}

   Consider  an equator $S^1$ of a standard two--sphere
   which contains points equally distant from both poles $N$ and $S$.
It is easy to see that this manifold is non-displaceable in $S^2$,
as any two great circles of a sphere do intersect. These
two mutually antipodal intersection points belong 
simultaneously to both great circles,
and their geodesic distances to the original poles $N,S$ of the sphere 
and to the rotated poles, $N'$ and $S'$, are equal
-- see Fig. \ref{fig:3nondis}.

The above statement can be formulated more precisely as

\smallskip
{\bf a)} a great circle $S^1$ embedded in $S^2$ is not displaceable
    with respect to $O(3)$ transformations.
\smallskip

This simple fact can be generalized  for higher dimensions in various ways
and concerns 
Lagrangian submanifolds of a space with a symplectic structure.

\smallskip
{\bf b)}  a Clifford torus $T^{M}$
  (flat embedding of the Cartesian product of $M$ great circles)
   embedded in a complex projective space $\mathbb{C}P^{M}$
   is non-displaceable with respect to transformations 
by a unitary $U\in {U}(M+1)$ -- see  \cite{Cho04,Tam08};
\smallskip

{\bf c)} the real projective space $\mathbb{R}P^{M}$
is non-displaceable in $\mathbb{C}P^{M}$  with respect to transformations
by ${U}(M+1)$ \cite{Oh93,Tam08}.
\smallskip

Observe that in analogy to case a) 
also in cases b)--c) the dimension $M$ of the
non-displaceable manifold is equal to half of the real dimension $2M$
of the embedding space.
These mathematical results find their direct applications
in quantum physics. Since a great torus $T^{M}$ 
is non-displaceable
in a complex projective space $\mathbb{C}P^{M}$,
for any choice of two fixed bases in ${\cal H}_{M+1}$
there exist quantum states {\sl mutually coherent},
as their coherences  with respect to both bases are maximal \cite{KJR14}.
 Analyzing the intersection points of any two Clifford tori $T^M$ 
 is interesting from the point of view of mutually unbiased bases
 and characterization of the set of unistochastic matrices  \cite{AB17}.
 
 As discussed in the previous section,
 for a two-qubit system,
the set  of maximally entangled states
reads, ${\cal P}_{\rm max}=U(2) /U(1)=SU(2)/Z_2$,
 where $Z_2$ denotes the finite group of order two. 
Making use of a 
homomorphism between classical groups, 
$SU(2)\sim SO(3)$,
one has  
${\cal P}_{\rm max}=O(3)/O(2)={\mathbbm R}P^{3}$.
As a real projective space embedded in the complex projective space 
is not displaceable -- see item c) above,
for any global unitary matrix $U\in U(4)$,
which determines the transformed partition $A'|B'$,
there exists a mutually entangled state,
maximally entangled with respect to both partitions \cite{PRCPZ15}. 

 The following generalization, 
based on numerical results, was conjectured in  \cite{PRCPZ15}:
for any two splittings $A|B$ and $A'|B'$ of the
$d \times d$ quantum system, related by an arbitrary $U\in U(d^2)$,
 there exist a {\sl mutually entangled} state,
maximally entangled with with respect to both partitions. 
From a geometric perspective this statement  is equivalent to

\smallskip
{\bf d)} manifold corresponding to projective unitary space,
 $PU(d)=U(d)/U(1)$, is {\sl conjectured to be} non-displaceable in $\mathbb{C}P^{d^2-1}$
   with respect to transformations by ${U}(d^2)$,
 
\smallskip
  which has not been proven, so far.
\smallskip 

\section{Multipartite systems and absolutely maximally entangled states}
\label{sec_AME}

Any quantum system consisting of $N$ subsystems
can be described in a Hilbert space
composed of $N$ factors, 
 ${\cal H}={\cal H}_A \otimes  {\cal H}_B \otimes \cdots \otimes {\cal H}_{N}$.
 Assuming, for simplicity, that all local dimensions are equal to $d$,
 the entire space has $d^N$ complex dimensions.
 Any pure state $|\psi\rangle \in {\cal H}$
 can be represented in a product basis by a tensor $T$
 with $N$ indices, each running from $1$ to $d$,
 \begin{equation}
|\psi\rangle = \! \sum_{i_1,\dots, i_N=1}^d \!\! T_{i_1, i_2, \dots i_N}
|i_1\rangle \otimes |i_2\rangle \otimes \cdots \otimes  |i_N\rangle.
\label{stateN}
\end{equation}

In analogy to the bi-partite case, such a state is separable
if the tensor $T$ is of rank one, 
so there exists a product basis in which the above sum reduces
to a single term only. In the bi-partite case, $N=2$
the tensor $T$ reduces to a matrix, denoted in Eq. (\ref{state2}) by $C$,
for which it is simple to find its rank $r$, its norm
and to perform its singular value decomposition.
In general, for a tensor with $N\ge 3$ indices,
there is no direct analogue of singular value decomposition 
\cite{LMV00a,CHS00,SA14},
and even evaluation of a rank of a tensor 
(there are various variants of this notion!)
becomes difficult  \cite{Kr77,St83,La12}.

To characterize the degree of entanglement quantitatively
one can apply geometric measures of entanglement,
initially developed for bipartite systems \cite{VP98,ZB02,WG03},
and generalize them for a multipartite scenario \cite{TWP09}.
In short, entanglement of a given state $|\psi\rangle$ 
can be defined by its distance $D_{\rm min}$ to the 
closest separable state $|\phi_{\rm sep}\rangle \in
 [\mathbb{C}P^{d-1}]^{\times N}$. 
Using the natural geodesic distance on the complex projective space,
equivalent to the Fubini-Study distance, one gets a simple expression, 
 $D_{\rm min}(\psi)=\arccos\bigl( |\langle \psi|\phi_{\rm sep}\rangle| \bigr)$,
 but the problem consists in identification of the closest
 separable state, which is often difficult. 
 
 In the bipartite case (\ref{state2}), the overlap
 is determined by the largest singular value
 of the matrix $C$ of coefficients,
 $|\langle \psi|\phi_{\rm sep}\rangle |=\sqrt{\lambda_{\rm max}}$,
 equal to the spectral norm $||C||_{\infty}$.
 For a multipartite system, this generalizes directly  \cite{BFZ19}
  to the  spectral norm  $||T||_{\infty}$ of a tensor \cite{FW20}.
  Alternatively, the degree of entanglement can be also characterized 
  by the nuclear norm $||T||_1$ of a tensor,  dual to the spectral norm,
  but evaluation of both quantities is hard \cite{FL18}. 
To overcome cumbersome problems with tensor algebra
one can investigate multipartite entanglement by
studying bi-partite entanglement for various splittings \cite{Sc04,FFPP08,FFMPP09}.

 An interesting class of highly entangled  $N$-partite states  
 is called {\sl $k$-uniform} \cite{Sc04,AC13}. 
 Such a state $|\psi_N\rangle \in {\cal H}_d^{\otimes N}$
  is distinguished by the following condition:
 for any choice of $k$ out of $N$ subsystems, 
 the reduced density matrix,
  obtained by partial trace over the remaining $N-k$ subsystems,
 is maximally mixed,
 \begin{equation}
 \rho_k={\rm Tr}_{N-k} |\psi_N\rangle \langle \psi_N|  \propto {\mathbbm I}_{d^k}.
\label{k_uniform}
\end{equation}
This implies that for any choice of $k$ subsystems
quantum correlations between them and the remaining $N-k$ subsystems are maximal. 
As the rank of a partial trace of an $d_1 \times d_2$ bipartite system
is not larger than min$(d_1,d_2)$, property 
(\ref{k_uniform}) can hold if the number $k$ of remaining subsystems
is not larger than $N-k$. Hence the maximal degree of uniformity of an
 $N$-party pure state is $k_{\rm max}=[N/2]$.
 Therefore such multipartite states are called 
{\sl absolutely maximally entangled}  (AME)  \cite{HCLRL12},
as they display extremal bi-partite correlations 
among all possible splittings. 

It is known that there are no such states for $N=4$ qubits \cite{HS00,HGS17},
but such states exist for five and six qubits.
Furthermore, by increasing the local dimension one can find AME states of four
qutrits --  subsystems with local dimension $d=3$,
first analyzed by Popescu,
\begin{eqnarray}
\label{Popescu}
|{\rm AME}(4,3)\rangle&=& |1111\rangle+|1223\rangle+|1332\rangle+ \nonumber\\
                        &&   |2122\rangle+|2231\rangle+|2313\rangle+\nonumber\\
                        &&  |3133\rangle+|3212\rangle+|3321\rangle,
\end{eqnarray}
where the normalization constant is omitted for brevity.
Allowing the indices to run from $0$ to $d-1$ 
the same state can be written in a more compact form,
\begin{equation}
|{\rm AME}(4,3)\rangle=\sum_{i,j=0}^2|i\rangle|j\rangle|i\oplus j\rangle|i\oplus 2j\rangle,
\label{popescu3}
\end{equation}
where addition $\oplus$  inside the kets is understood modulo $d=3$.
It is easy to check that this state possesses the required properties
as it is $k=2=N/2$ uniform -- all partial traces of the projector
 over any two subsystems are maximally mixed,
so the quantum correlations between all four subsystems are maximal.

Further examples for $d=4,5$ and $d\ge 7$ 
can be generated 
from classical combinatorial designs called mutually orthogonal Latin squares
as described in Section \ref{sec_classOLS}.  
A list of dimensions $d$ and numbers of subsystems $N$,
for which states AME$(N,d)$ exit  \cite{HESG18}
 is also available online \cite{HW_table}.

Let us concentrate here on the case $N=4$,
such that any pure state $|\psi\rangle$ is represented in 
(\ref{stateN}) by a tensor $T_{ijk\ell}$ with four indices,
each running from $1$ to $d$,
which are related to four subsystems,  $ABCD$.
 Three possible symmetric splittings of the system,
 $AB|CD$, $AC|BD$ and $AD|BC$, 
 correspond to three different flattenings
of the tensor $T_{ijk\ell}$ into matrices $T_{\mu\nu}$ of order $d^2$,
where both composed indices, $\mu,\nu=1,\dots, d^2$,
are obtained by grouping together  two out of four initial indices $i,j,k,\ell$.

\begin{figure}[h] 
\begin{center}
\includegraphics[width=0.26\textwidth]{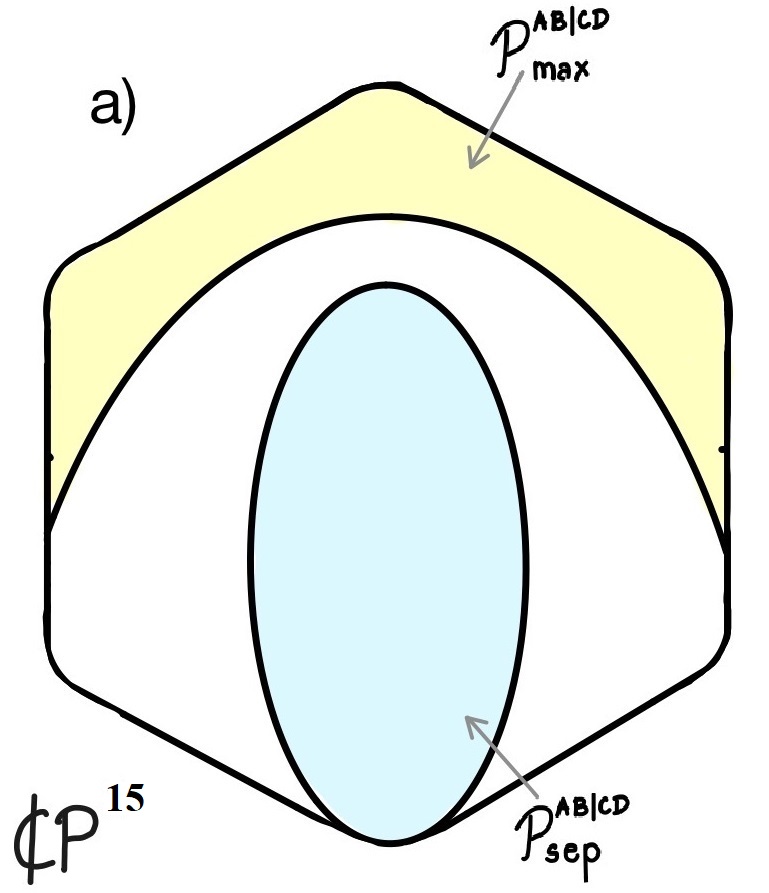}
\includegraphics[width=0.30\textwidth]{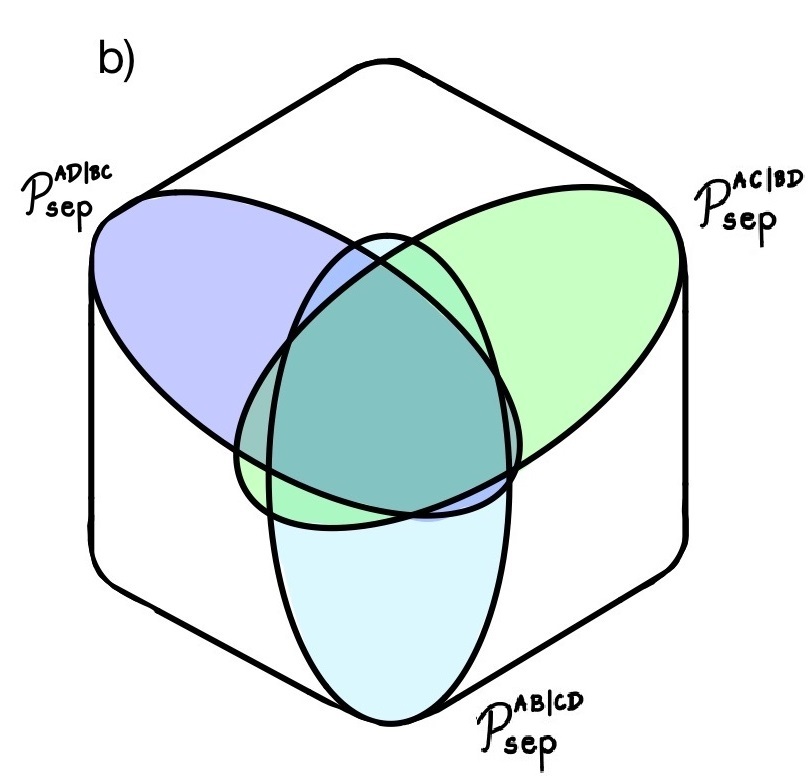}
\includegraphics[width=0.30\textwidth]{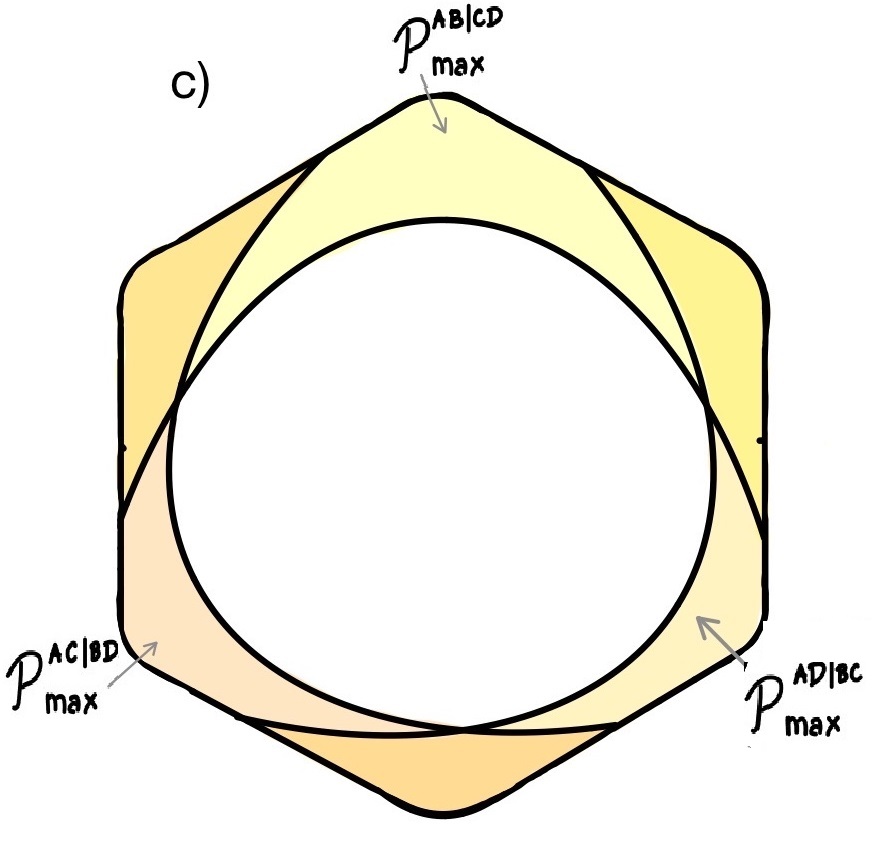}
\caption{Four-qubit system $ABCD$ can be treated as a
bipartite, two-ququart systems in three different ways:
a) shows the sets of separable 
${\cal P}^{AB|CD}_{\rm sep}\subset {\cal P}_{16}$
and maximally entangled states,  ${\cal P}^{AB|CD}_{\rm max}$,
for a selected splitting;  states separable with respect to  
three different splittings are shown in panel b),
while panel c) presents analogous sketch for maximally entangled states
${\cal P}_{\rm max}=PU(4)=U(4)/U(1)$. 
Note that any two sets  ${\cal P}_{\rm max}$ do intersect, 
but the intersection of all three of them is empty.}
\label{fig:4qubit}
\end{center}
\end{figure}

As discussed in Section  \ref{sec_separable}
maximal entanglement of any  bi-partite, $d \times d$ state
requires that the matrix of coefficients is unitary, up to rescaling
$CC^{\dagger}={\mathbbm I}/d$.
Thus a $4$-party AME state, satisfies simultaneously unitarity conditions
for three different splittings:  the matrices 
$X_{\mu\nu}=T_{ij,k\ell}$,
$Y_{\mu\nu}=T_{ik,j\ell}$,
$Z_{\mu\nu}=T_{i\ell,kj}$,
with index $\mu$  determined by  a)  pair $(i,j)$;  b) pair  $(i,k)$,  and c) 
pair $(i,\ell)$, are unitary up to a constant.
 A  four-index tensor $T_{ijk\ell}$ with 
the property that its  three flattenings are proportional to a unitary 
is called {\sl perfect}  \cite{PYHP15}. 
The resulting unitary matrix $U=Xd$ of size $d^2$
with the property that its both reorderings $Yd$ and $Zd$ are also unitary,
 is called  {\sl two-unitary} \cite{GALR15}. These reorderings
 of entries of the matrix, induced by permutations of indices of 
 $dT_{ijk\ell}$, 
 are often called reshuffling and partial transpose,
 written as $U^R=Yd$ and $U^{\Gamma}=Zd$ respectively, 
 see  \cite{ZB04,RBBRLZ22}.

Observe that a change of the partition $AB|CD$ into $AC|BD$,
 corresponds to a different choice of a pair of indices of
the tensor $T_{ijk\ell}$, and also to a unitary permutation of 
order $d^4$, which induces a transformation of the 
subspace ${\cal P}_{\rm max}$ of maximally entangled states of the
bi-partite, $d^2\times d^2$ system. It is thus natural
to refer to the theory of non-displaceable manifolds,
discussed in Sec. \ref{sec_nondisp}. The conjectured property d)
implies that arbitrary `two copies' of the manifold 
${\cal P}_{\rm max}=PU(d^2)$  do intersect in the embedding space 
 ${\cal P}_{d^4}$. However, this fact does not imply the AME property, which
  requires {\sl three} concrete `copies' of this manifold, 
  corresponding to three different splittings of the system,
  do intersect in a single point.
 
The work by Higuchi and Sudbery \cite{HS00} implies
that the dimension  $d=2$ is `not big enough'
to allow for the existence of AME states of four qubits. 
Let us illustrate this fact in terms of intersections of manifolds of
maximally entangled states for this $4$-party system,
 treated here as a bi-partite,  $d^2\times d^2$ system.  

In principle, one could follow the scheme of \cite{EHMS21},
and in analogy to Fig.~\ref{fig:2qutrit}, 
consider a tetrahedron of Schmidt coefficients of a
$4\times 4$ pure states, plot the orbit of separable states,
    ${\cal P}_{\rm sep}={\mathbbm C}P^{3} \times {\mathbbm C}P^{3}$,
    which connects four corners of the simplex,
       and place  the manifold
         ${\cal P}_{\rm max}= PU(4)$ of maximally entangled states,
         at the center of the tetrahedron, $\Lambda_*=\frac{1}{4}(1,1,1,1)$.
   However, as we need to consider three different splittings of a four-party
   system $ABCD$,  Fig. \ref{fig:4qubit} is constructed in analogy to           
   Fig. \ref{fig:2qubit} representing pure states for a two-qubit system.

In Fig. \ref{fig:4qubit}a) we show the set of states which are separable,
${\cal P}^{AB|CD}_{\rm sep}$,
and maximally entangled,  ${\cal P}^{AB|CD}_{\rm max}$,
with respect to this splitting  $AB|CD$ of the system.
Both sets are located at the distance $D=\arccos(1/2)=\pi/3$ apart.
Further panels show analogous sets for other splittings.
Three copies of the set   ${\cal P}_{\rm sep}$
do intersect in the center of
Fig. \ref{fig:4qubit}b),
as any fully separable state, 
$|\phi_A\rangle \otimes 
|\phi_B\rangle \otimes 
|\phi_C\rangle \otimes 
|\phi_D\rangle$,
  is manifestly separable with respect to all three splittings.
As the set   ${\cal P}_{\rm max}=PU(4)$
 is conjectured to be non-displaceable in ${\cal P}_{16}$
  two copies of this set do intersect.
However, the joint intersection of three such copies is empty
 as sketched in  Fig. \ref{fig:4qubit}c), 
 since there are no AME$(4,2)$ states.

\section{Classical orthogonal \hskip 3.0cm
 Latin squares}
\label{sec_classOLS}

A classical combinatorial design consists of discrete symbols 
arranged  with a particular `symmetry and balance' \cite{CD07}. 
 A simple example
 is provided by a {\sl Latin square} of order $d$,
 which consists  $d$ copies of $d$ symbols,
 arranged into a square of size $d$ in such a way, that each row and each 
 column contains different symbols. It is easy to see that such configurations exist
 for any dimension $d$.
 
\begin{figure}[htbp]
\begin{center}
\includegraphics[width=0.42\textwidth]{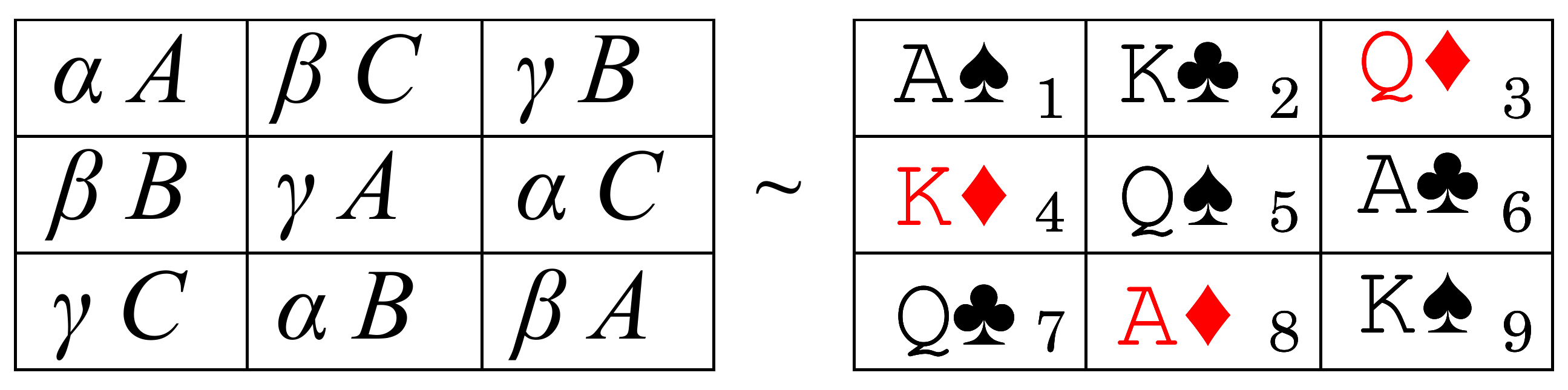}
\caption{Two orthogonal Latin squares of order three, also called  
Graeco-Latin square, can be visualized with a set of nine cards,
labeled from $1$ to $9$.
}
\label{fig:GL3}
\end{center}
\end{figure}

 Leonhard Euler analyzed such designs and became interested 
 in constructing pairs of Latin squares that satisfy an additional condition of
 {\sl orthogonality}: all possible pairs of $d^2$ symbols are present in the square.
 A pair of orthogonal Latin squares is also called a {\sl Graeco-Latin square}, 
 since Euler represented one square with Greek letters and the other one
 with Latin. A simple example of such a pattern, written
 OLS(3), i.e. orthogonal Latin squares of order three, is shown
 in Fig.\ref{fig:GL3}.

Orthogonality of both squares implies that all pairs of symbols (nine cards in the figure)
 are different.
Thus  the pattern 
shown in Fig.~\ref{fig:GL3}
can be encoded by the permutation matrix $P_9$
  of order $9$. It determines which card, 
ordered in a way natural for bridge players,
 $\{ \text{A\spade}, \text{A\diamond}, \text{A\club}, 
  \text{K\spade},  \text{K\diamond}, \dots, \text{Q\club} \}$,
should be placed in which position of the square, 
labeled in the bottom line,
\begin{equation}
		\label{P9}
		\!\!
		P_9=
		\begin{array}{c}
			\left(\begin{array}{ccc|ccc|ccc}
				\text{A\spade} &0&0&0&0&0&0&0&0\\
				0&0&0&0&0&0&0&\text{A\diamond}&0\\
				0&0&0&0&0&\text{A\club}&0&0&0\\
				\hline
				0&0&0&0&0&0&0&0&\text{K\spade}\\
				0&0&0&\text{K\diamond}&0&0&0&0&0\\
				0&\text{K\club}&0&0&0&0&0&0&0\\
				\hline
				0&0&0&0&\text{Q\spade}&0&0&0&0\\
				0&0&\text{Q\diamond}&0&0&0&0&0&0\\
				0&0&0&0&0&0&\text{Q\club}&0&0
			\end{array}\right) \\
			\begin{array}{ccccccccc}
				{\text{{\color{white}A\!\!\!\!\!\spade}}}_{\!\!\!\!\!1}&
				{\text{{\color{white}A\!\!\club}}}_{\!2}&
				{\text{{\color{white}A\!\!\wdiamond}}}_{\!3}&
				{\text{{\color{white}A\!\!\wdiamond}}}_{4}&
				{\text{{\color{white}A\!\!\spade}}}_{\!5}&
				{\text{{\color{white}A\!\!\club}}}_{\!6}&
				{\text{{\color{white}A\!\!\club}}}_{7}&
				{\text{{\color{white}A\!\!\wdiamond}}}_{\!8}&
				{\text{{\color{white}A\!\!\spade}}}_{\!9} 
			\end{array}
		\end{array},
	\end{equation}
	where each card represents the number $1$. 
 To assure that each rank and each suit of cards do not repeat in each row
  and column of the square the  non-zero entries of the matrix $P_9$ satisfy the rules
of a strong Sudoku: in each row, column and block there is a single
non-zero entry. Moreover, all the
locations of these entries in each block are different.
Alternatively, the design can be described by a table of size $9 \times 4$, 
in which for each card we write four digits: the value `v' (or rank of a card), its suit `s', the row `r'
and the column `c' in the square, each running from $1$ to $3$. 
Conditions of OLS imply that there exist three {\sl invertible} functions, 
which map a pair of two digits into the other two, $(v, s) = F_1(r, c)$; \;
$(v, r) = F_2(s, c)$ and $(v,c) = F_3(s, r)$.

In other words, the permutation matrix $P_9$ is two-unitary, and restructured into 
$T_{ij,k\ell}$ forms a perfect tensor with four indices running from $1$ to $3$,
so that the partially transposed matrix $P_9^{\Gamma}$ and 
the reshuffled matrix $P_9^R$ also form permutations, and are thus unitary and invertible.
Making use of the coding:  (A; K; Q) $ \leftrightarrow (1; 2; 3)$ and 
$(\text{\spade}; \text{\diamond} ; \text{\club} ) \leftrightarrow (1; 2; 3)$
one can represent the Graeco-Latin square in the 
numeric form,
\begin{equation} \
{\huge 
\begin{array}{|c|c|c|} \hline 
{ 1} \, {1}   & {2}\, {3}&  { 3} \, { 2}  \\ \hline
{ 2}\, {2}  & { 3} \, {1} &  { 1}\, { 3} \\  \hline
 {3}\, {3}  & { 1}\, {2}&  { 2} \, {1}  \\  \hline
\end{array} 
}
\label{GL3}
\end{equation}

\begin{figure}[htbp]
\begin{center}
\includegraphics[width=0.35\textwidth]{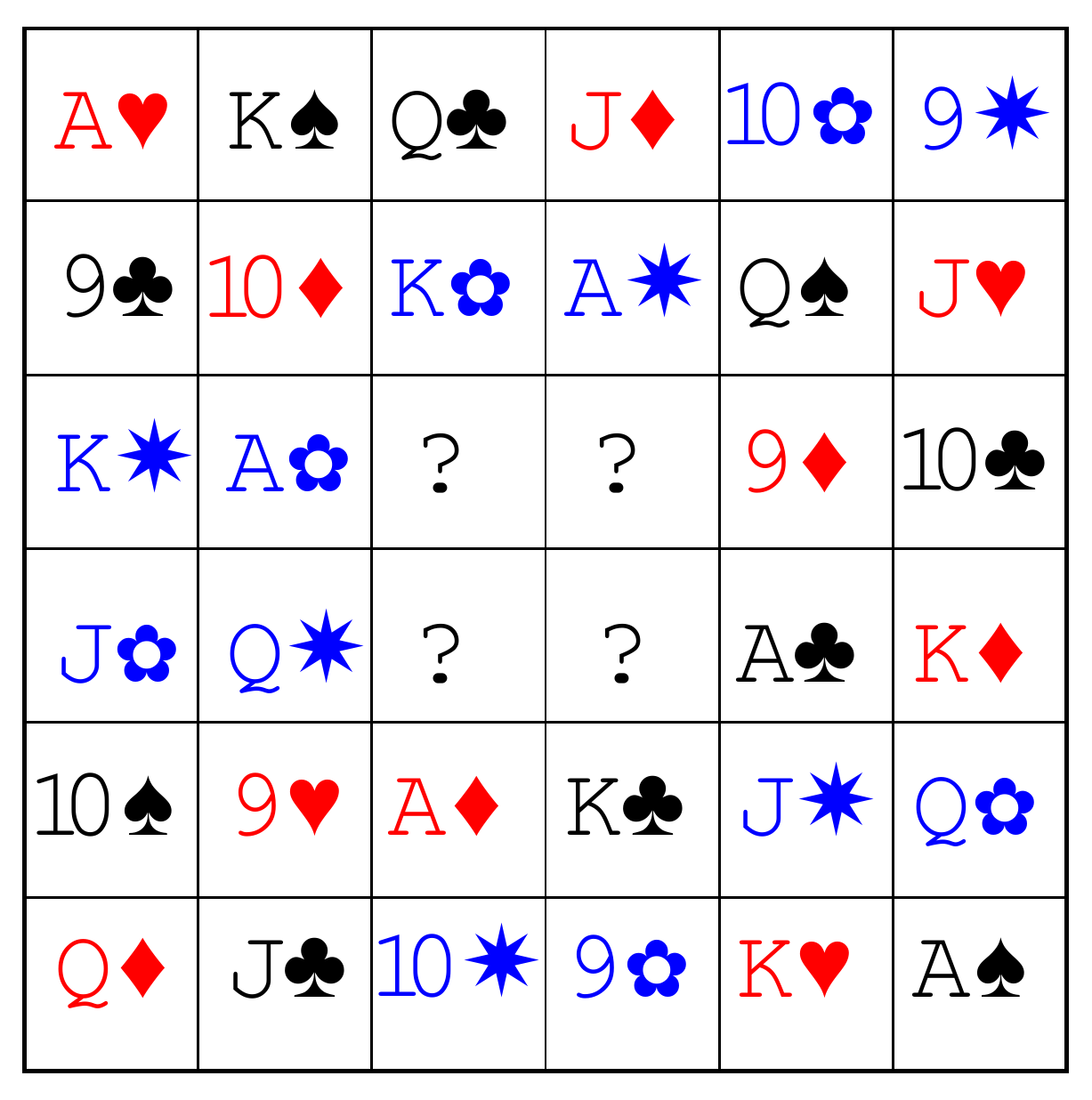}
\caption{To find a solution of the Euler problem of $36$ officers
one needs to place four missing cards into the center of the square,
in such a way that neither the ranks of the cards
 nor their suits are repeated in each row and each column.}
\label{fig:OLS6}
\end{center}
\end{figure}

To see a direct relation to quantum entanglement
observe that the above design allows us to construct
the state AME$(4,3)$ appearing in Eq. (\ref{Popescu}):
each of the nine terms of the four-partite state corresponds to a single card,
two first digits are given by its position in the square (row and column),
the third one is determined by its rank and the fourth by the suit.
As this construction works for an OLS of an arbitrary dimension,
we conclude that the states AME$(4,d)$ exist for all these dimensions,
for which OLS$(d)$ are known. It is easy to construct such a design if $d$ is a prime 
or a power of a prime. Furthermore, OLS are known to exist 
for any $d\ge 7$ \cite{JCD01}. However, such a configuration 
does not exist  for $d=6=2\times 3$,
as conjectured by Euler, who formulated his famous problem of $36$ officers,
coming from  six different units,  of six ranks in each unit.
This fact was proved in 1900 by Tarry \cite{Tarry,Stin84}.
To put some light on Euler's problem we invite the reader 
to study a particular attempt to solve it, 
shown in Fig.~\ref{fig:OLS6}.

 As there is no solution of the original problem of Euler, there are no $2$-unitary permutation
 matrices of size $36$.
  However, one can relax
 the assumption that we look for a permutation matrix of order $36$ 
 and look for a solution in the larger continuous group of unitary matrices $U(36)$.
 This step is equivalent to leaving the well studied ground of classical 
 combinatorial designs, and moving to a more general  
 class of their quantum analogues.
 
\section{Quantum orthogonal \hskip 3.0cm
 Latin squares}
\label{sec_quantOLS}

As a known rule of thumb, every good notion can be quantized.
A cornerstone of the flourishing new field of {\sl quantum combinatorics}
was laid by Gerhard Zauner, who introduced quantum analogues of classical designs 
in his  Ph.D Thesis from 1999 \cite{Za99}.
A quantum combinatorial design consists of quantum states from
a complex Hilbert space ${\cal H}_d$, arranged with 
a certain  `symmetry and balance'.

Such an approach to the classical notion of Latin squares was first 
advocated by Musto and Vicary \cite{MV16}, who proposed to
replace classical symbols in the square by pure quantum states.
They  defined a {\sl quantum Latin square} (QLS) of size $d$,
which consists of $d^2$ quantum states
$|\psi_{ij}\rangle \in {\cal H}_d$, 
 with $i,j=1,\dots, d$,
such that each column and each row of the square form an orthogonal basis.
The classical condition: all objects in each row and each
column are {\sl different}, is replaced here by its natural quantum analog:
all quantum states in each row and column are {\sl orthogonal}.

\begin{figure}[htbp]
\begin{center}
\includegraphics[width=0.49\textwidth]{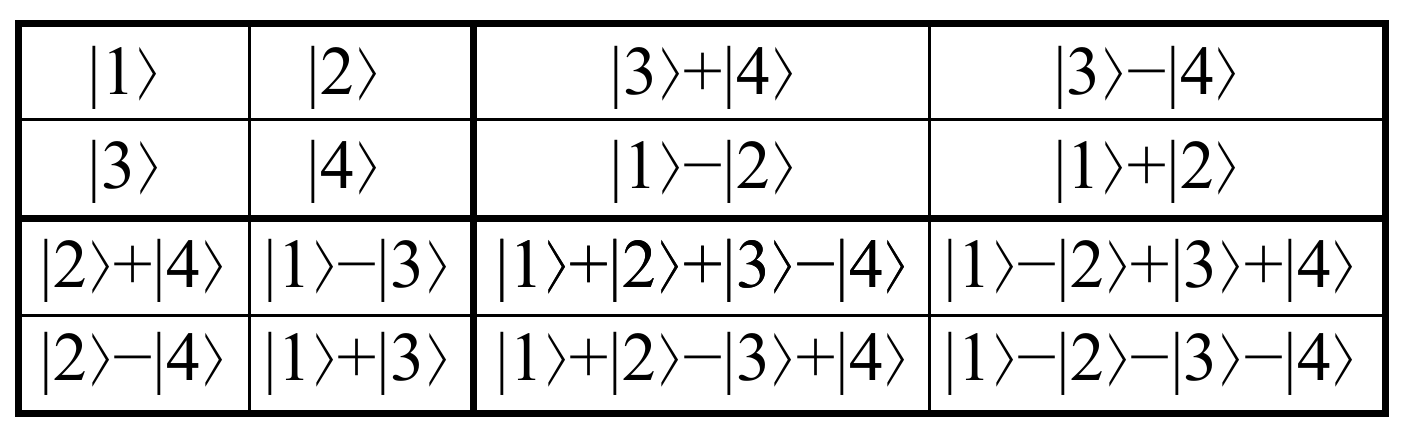}
\caption{Quantum Latin square of order $d=4$,
 which consists of $16$ different states -- normalization
 constants are omitted. Each of its rows and  columns
forms an orthogonal basis in ${\cal H}_4$,
as required. This design satisfies also a stronger condition
of a quantum sudoku, as four states in each $2 \times 2$ block also 
form an orthogonal basis.}
\label{fig:sudoku}
\end{center}
\end{figure}

An example of a quantum Latin square shown in 
Fig.~\ref{fig:sudoku}, borrowed from \cite{PWRBZ21},
satisfies also rules of a {\sl quantum sudoku} \cite{NP21}:
each row, column and  block form an orthogonal basis.
Note that any classical Latin square (or a classical sudoku pattern)
can be `quantized' in a trivial way, by placing each symbol
inside a ket and treating it as a quantum state $|\psi\rangle$.
In such a case the design contains $d$ copies of $d$
different states which form several copies of the standard computational basis. 
On the other hand, the quantum design shown in Fig. \ref{fig:sudoku}
consists of $16$ different states $|\psi_{ij}\rangle$,
which form together $12$ different orthogonal bases.
Replacing each state $|\psi_{ij}\rangle$ in field $(i,j)$ of the square
by the projector $|\psi_{ij}\rangle \langle\psi_{ij}|$
one obtains a {\sl blockwise bistochastic} matrix: positive operators in each row and
each column sum to identity. Such designs were recently analyzed in 
\cite{CDN20,CNV22} under the name {\sl quantum magic squares}.
The design presented in  Fig.~\ref{fig:sudoku} certainly
deserves such a name, as all its entries are different, just as in the
classical case.

To look for a broader class of quantum AME states
one has to consider quantum analogues of  two orthogonal Latin squares.
The first attempt to construct such a quantum pattern \cite{GRMZ18},
was later improved in \cite{MV19},
but we follow here the version used in \cite{Ri20,RBBRLZ22}.
Before presenting a definition of such a  quantum combinatorial design, 
let us list the rules of Euler defining a classical OLS(d), which consists of $d^2$ symbols,
each described by its color and rank:
\smallskip

{\bf  C1)}  all $d^2$  symbols  are different,
\smallskip

{\bf C2)} all $d$ symbols in each row are of different colors and different ranks,
\smallskip
 
{\bf C3)} all $d$ symbols in each column are of different colors and different ranks.

\medskip

This list will be helpful to understand the corresponding quantum 
constraints.

\medskip 

{\bf Definition.}
 {\sl A pair of  {\sl quantum  orthogonal Latin squares} (QOLS) 
of size $d$  is defined as a set of $d^2$ bipartite states,
$|\psi_{ij}\rangle \in {\cal H}_A \otimes {\cal H}_B$
which form a square,}
\begin{equation} \
{\large 
{\rm QOLS} \; :=  \;
\begin{array} {|c c c|} \hline 
|\psi_{11}\rangle & \dots & |\psi_{1d}\rangle \\
      \dots            & \dots  &    \dots     \\
|\psi_{d1}\rangle & \dots & |\psi_{dd}\rangle \\
\hline
\end{array} \; ,
}
\label{QOLS2}
 \end{equation}
{\sl and satisfy  constraints  {\bf Q1)}--{\bf Q3)} specified below,
analogous to the classical conditions {\bf C1)}--{\bf C3)}.} 
\medskip

There is a natural way to quantize condition {\bf C1)}, 
 imposing that all the objects of the classical design are different:

\medskip
{\bf Q1)} All $d^2$ states in the pattern are mutually orthogonal 
and they form a basis
  in ${\cal H}_d \otimes {\cal H}_d$,
\begin{equation}
\label{a_prim}
\langle \psi_{ij}|\psi_{k\ell}\rangle \; =\;  \delta_{ik} \delta_{j \ell}.
\end{equation}

\medskip

The classical conditions {\bf C2)}--{\bf C3)} require some explanation.
In each row (or column) we
analyze separately colors (or ranks) of the symbols in the classical pattern. 
Thus one can expect that the analogous quantum constraints 
concern the partial traces of symmetric combinations of projectors.
As in the classical design no color is repeated in each row
and column the `average color'  should be `white'. 
Therefore, in quantum designs, it is assumed \cite{GRM21} 
that the partial trace over the second subsystem $B$ (or the first subsystem $A$)
is maximally mixed and there are
no correlations between different rows (and different columns).

\medskip
{\bf Q2)} All rows of the square satisfy the partial trace conditions
\begin{equation}
\label{b_prim}
   \sum_{k=1}^{d} {\rm Tr}_B \bigl(  |\psi_{ik}\rangle \langle \psi_{jk}|  \bigr) \; =\; \delta_{ij}
{\mathbbm I}_d\;.
\end{equation}
Dual conditions hold for the other partial trace ${\rm Tr}_A$,
 analogous to the requirement that all ranks in each row of the classical design are different.
 
\medskip

{\bf Q3)} All columns of the square satisfy analogous conditions
\begin{equation}
\label{c_prim}
 \sum_{k=1}^{d}  {\rm Tr}_B   \bigl( |\psi_{ki}\rangle \langle \psi_{kj}| \bigr) \;= \; \delta_{ij} 
{\mathbbm I}_d \; ,
\end{equation}
 \hskip 1.3cm  and dual conditions for ${\rm Tr}_A$.
\medskip

Observe that the above conditions are satisfied
by any classical design, upgraded to a quantum setup
by replacing any symbol by the corresponding product state,
e.g.  
$\text{A\spade} \to  |\text{A\spade}\rangle =|11\rangle \in {\cal H}_d\otimes {\cal H}_d$.
Since conditions {\bf Q1)- Q3)} are invariant with respect to 
local unitary operations, $U_A \otimes U_B$, with $U_A, U_B\in U(d)$, 
 they are also satisfied by  orthogonal apparently quantum Latin squares,
obtained from any classical solution by a local unitary transformation.
However, a legitimate quantum solution can contain 
also entangled states  $|\psi_{ij}\rangle$, for instance a Bell state,
$|\text{A\spade} \rangle + |\text{K\diamond}\rangle$,
also written as $|11\rangle + |22\rangle$,  
so in general
it is not possible to split a pair of QOLS into two separate classical designs OLS.

Note that the above three requirements  
ensure that the  state generated from the collection of $d^2$
bipartite states, $\{|\psi_{ij}\rangle\}_{i,j=1}^d$ forming a QOLS, 
\begin{equation}
\label{AME4}
|\Psi_{ABCD}\rangle = \frac{1}{d} \sum_{i,j=1
}^{d}  |i\rangle_A |j\rangle_B |\psi_{ij}\rangle_{CD}, 
\end{equation}
is an  AME$(4,d)$ state \cite{RBBRLZ22}.
To show this let us combine indices $(i,j)$ into a composite index $\mu=1\dots d^2$
and merge  the vectors $|\psi_{\mu}\rangle$
into a matrix  $U_{\mu,\nu}$,
 where the other composite index, $\nu \leftrightarrow (k,\ell)$,
labels $d^2$ components of each vector $|\psi_{\mu}\rangle$.
Then the quantum conditions  {\bf Q1)-Q3)},
related to the maximal entanglement for three different splittings of the system $ABCD$,
 imply that the matrix $U_{\mu,\nu}$ is $2$-unitary,
or equivalently, 
the corresponding tensor $T_{ijk\ell}$ is perfect,
as necessary and  sufficient to generate a four-party AME state \cite{GALR15}.

In this way we arrived at the following statement \cite{Ri20,GRM21}. 

\medskip
{\bf  Observation 1.} 
{\sl Existence of a state $|{\rm AME}(4,d)\rangle \in {\cal H}_d^{\otimes 4}$
is equivalent to:
\smallskip

E1) existence of {\rm QOLS}(d);
\smallskip

E2) existence of a $2$-unitary matrix $U\in U(d^2)$;
\smallskip

E3) existence of a perfect tensor $ T_{ijk\ell}$ 
  with four indices, each running from $1$  to $d$.
  }
\medskip

A perfect tensor $T$ reshaped into rows, 
$i, j, |\psi_{ij}\rangle$ forms a 
quantum orthogonal array (QOA)
with $4$ columns,  $d^2$ rows
of strength $k=2$ and alphabet containing of $d$ symbols \cite{GRMZ18}.
The array can be divided into two classical columns $i,j$,
 and two quantum columns,  containing possibly entangled states $|\psi_{ij}\rangle$,
 so it  is written QOA$(d^2, 2_C+2_Q,d,2)$. 
An AME state of four subsystems with $d$  levels each
leads to a quantum code $((4, 1, 3))_d$,
which allows one to encode one state of a $d$-level systems in four such systems
and correct a single error \cite{GBR04}. 
Therefore, Observation 1 concerning equivalence relations
can be extended  by the following two items:

\medskip
{\sl
E4) existence of a quantum orthogonal array of strength $k=2$
with $d^2$ rows, $4$ columns and $d$-letters symbols,
denoted as {\rm QOA}$(d^2, 2_C+2_Q,d,2)$. 

\smallskip
E5) existence of a four-qudit quantum code  $((4, 1, 3))_d$.}
\medskip
   
Before discussing the Euler's case $d=6$
let us return to the simpler case of $d=2$,
for which there are no Graeco-Latin squares.
As discussed in \cite{ZBRBRL22}
even using two-qubit entangled states
it is not possible to construct a QOLS of order two,
in accordance to the result of Higuchi and Sudbery, 
who demonstrated that there are no AME states for a
four-qubit system  \cite{HS00}.

\section{Solution of the quantum Euler problem for $d=6$ and its geometric consequences}
\label{sec_deq6}

The problem whether there exist quantum orthogonal Latin squares of order six,
or equivalently, AME state of four subsystems with six levels each,
was studied  for a few recent years \cite{GRMZ18,Ri20,YSWNG21},
but the solution was found only recently  \cite{RBBRLZ22}.
We looked for a $2$-unitary matrix $U\in U(36)$,
such that its partial trace $U^{\Gamma}$
and reshuffling $U^R$, are also unitary, 
so that its rows (or columns) 
treated as  pure states of a $6 \times 6$ system
determine the solution  $\{|\psi_{ij}\rangle\}_{i,j=1}^d$.

This task was performed numerically by the following iterative procedure.
For any initial matrix $X_0$ of size $36$ we take the corresponding 
unitary $V$ obtained by its polar decomposition, $X_0=HV$,
then write its reshuffling in the similar form, $X'=V^R=H'W$ with a unitary $W$.
The output matrix $X_1$ is then obtained by partial transpose of the resulting unitary, 
$X_1=W^{\Gamma}$. Repeating such a Sinkhorn-like algorithm \cite{ARL21}
several times one generates a sequence of matrices $X_n$,
but its convergence to a $2$-unitary matrix is not guaranteed. 
As the search is performed in the huge space $SU(36)$ of  $1295$ real dimensions,
the results depend critically on the choice of the seed $X_0$.
 
By inserting four cards missing in Fig.  \ref{fig:OLS6}
we obtain an approximation of the non-existing classical pattern OLS$(6)$
and generate a certain permutation matrix $P_{36}$,
which determines which card goes into which field of the square.
Such a matrix, maximizing the entangling power among all 
bi-partite permutations \cite{CGSS05},
if perturbed by a small random perturbation $Y$,
gives a seed of the type $X_0=P_{36}+ \epsilon Y$.
Such an initial point generates a trajectory $X_n$, typically
converging (at $n \sim 400$) to a matrix $U_{36}$,
two-unitary up to the numerical accuracy  \cite{RBBRLZ22}.
Making use of the freedom of local unitary transformations
it was possible to find a product basis,
 in which  $U_{36}$ becomes sparse  \cite{WB_thesis,AB_thesis},
 and to bring it to the block diagonal form. 
 
\begin{figure}[htbp]
\begin{center}
\includegraphics[width=0.47\textwidth]{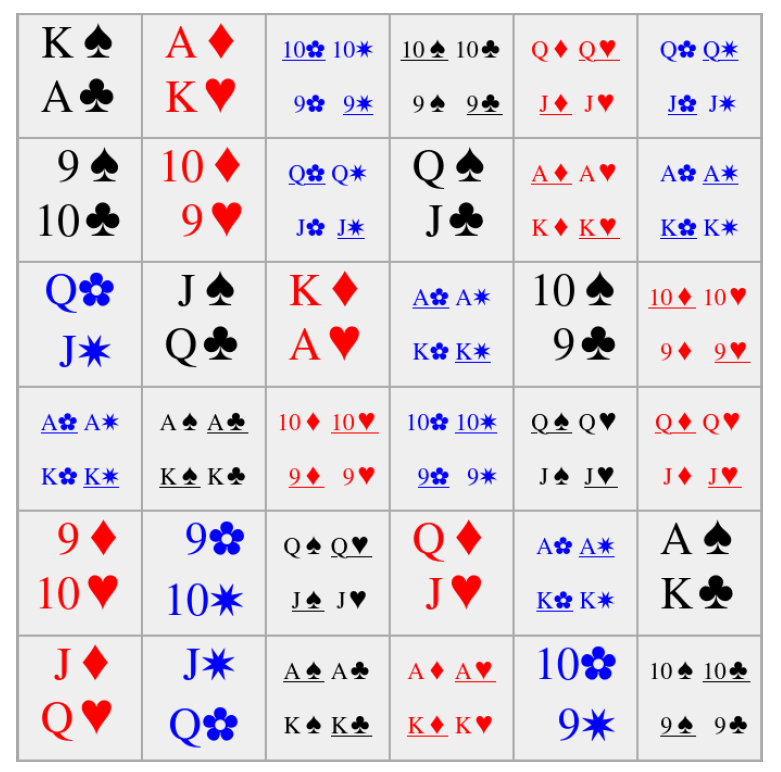}
\caption{Solution of the quantum version of the Euler problem
constructed in \cite{RBBRLZ22}: several cards in a single field 
represent an entangled state, 
say a Bell state,
$|\psi_{11} \rangle= |\texttt{K} \text{\spade}\rangle + |\texttt{A} \text{\club}\rangle$,
or in 
a usual notation, 
$|11\rangle+|22\rangle$, in the upper left corner.
 The size of each card encodes the amplitude in the superposition:
  large $c$ (large symbol), medium $b$ (smaller symbol), 
  and the smallest $a$ (underlined symbol),
 while analytically known complex phases are listed in the original work.
 This solution was called {\sl golden}, as the ratio $b/a$
 gives the golden mean. }
\label{fig:QOLS6}
\end{center}
\end{figure}

 Imposing unitarity in each of its nine blocks of size four
 one could derive a fully analytical solution of the problem.
 The resulting $2$-unitary matrix $U_{36}$
 has $112$ non-zero entries, not more than four in each row.
  All entries share one of three possible amplitudes,
  $a=\frac{1}{2}\sqrt{1-1/\sqrt{5}}$,  $b=\frac{1}{2}\sqrt{1+1/\sqrt{5}}$,
  and the largest one,  $c=1/\sqrt{2}$. Interestingly, the ratio
   of the two smaller numbers gives the golden mean, 
   $\varphi=b/a=(1+\sqrt{5})/2$, so the pattern was called
  a {\sl golden AME state}  \cite{RBBRLZ22}. The solution is visualized 
  in  Fig. \ref{fig:QOLS6} by $112$ cards of six suits and six ranks,
  placed in the square of size $6$, so that each field represents a single state $|\psi_{ij}\rangle$
  forming the QOLS$(6)$. There are no more than $4$ cards in each field
  and the size of each symbol represents its amplitude.
  Complex phases of each entry, being multiples of $\omega=\exp( i \pi/10)$,
  are provided in \cite{RBBRLZ22}. 
    
  Observe that  $36$ cards can be divided into nine groups of four cards
  of the same colour and neighbouring ranks, so that
  all entries of the quantum design, represented by cards in a single cell of the square,
   contain superpositions of the cards from a single group.  Looking at 
  Fig. \ref{fig:QOLS6} it is easy to realize that the average suits and ranks in each column
    and each row of the square are balanced. However, to  verify that    
  all  the  necessary conditions {\bf Q1)-Q3)} are indeed satisfied
    one needs to use the complex phases found, as explicitly shown in Ref. \cite{ZBRBRL22}.

We shall now analyze certain geometric consequence of the established
solution of the quantum version of the problem of Euler.
Consider, in general, an $N=4$ party system $ABCD$ with local dimension $d$.
The known fact that  for $d=2$ there are no AME states \cite{HS00},
implies the following statement concerning the manifolds
of maximally entangled states
${\cal P}_{\rm max}=PU(4)=U(4)/U(1)$
of the $4 \times 4 $ bi-partite system
determined by three symmetric splittings of $ABCD$.

\medskip
{\bf  Observation 2.} 
{\sl Non-existence of the state $|{\rm AME}(4,2)\rangle \in {\cal H}_2^{\otimes 4}$
implies that  the following intersection of three projective group manifolds,
corresponding to various splittings, 
all embedded in  ${\mathbbm C}P^{15}$,
is  empty,}
\begin{equation}
  PU(4)^{AB}_{CD}
  \cap  PU(4)^{AC}_{BD}  \cap  PU(4)^{AD}_{BC} =
    \emptyset .
\label{inter2}
\end{equation}

Let us emphasize at this point that conjecture {\rm d)} formulated 
in Section \ref{sec_nondisp}
concerning non-displaceability of the  projective group manifold $PU(4)$ 
in ${\mathbbm C}P^{15}$
implies that {\sl two} copies of this manifold do intersect,
but this does not imply that three copies will meet in a single point.
Unitary matrices $U\in U(d^2)$
with the property that the reshuffled matrix is unitary,  $U^R\in U(d^2)$, 
are called {\sl dual unitary} \cite{BKP19,BKP20}.
It comes then not as a great surprise that dual unitary matrices exist for any 
dimension \cite{RAL2020,CL21},
what is directly implied by the non-displaceability conjecture.
The same statement holds for unitary matrices,
for which partial transpose is unitary $U^{\Gamma} \in U(d^2)$,
analyzed earlier in \cite{BN17}.

In some analogy to properties of physical systems containing interacting spins,
the non-existence of AME states for a four-qubit system
 was  dubbed {\sl frustration} \cite{FFPP08},
as the requirement of maximal entanglement with respect to two given splittings
contradicts an analogous condition for the third splitting. 
As the number of constraints for an AME state is so big, that they
can not be simultaneously fulfilled for $d=2$,
the frustration disappears in higher dimensions.

The case $d=6$ was particularly interesting, as
the lack of solution of the classical Euler problem of $36$ officers
did not allow one to generate the corresponding four-quhex entangled state.
The explicit analytic solution of the quantum version of the problem
constructed in \cite{RBBRLZ22} has also a geometric  interpretation.

\medskip
{\bf  Observation 3.} 
{\sl Existence of the state $|{\rm AME}(4,6)\rangle \in {\cal H}_6^{\otimes 4}$
implies that  the intersection of the following three manifolds  
embedded in  ${\mathbbm C}P^{36^2-1}$ is non-empty,}

\begin{equation}
 PU(36)^{AB}_{CD}
  \cap  PU(36)^{AC}_{BD}  \cap  PU(36)^{AD}_{BC}
  \supset  \{ {\rm AME}(4,6) \}.
\label{inter6}
\end{equation}

It is rather difficult to imagine how these high-dimensional manifolds 
embedded in a complex projective space do intersect.
Thus,  in Fig. \ref{fig:three_circles} 
we use the standard sphere $S^2$ to
present a simple caricature of this property.

\section{Concluding remarks}
\label{sec_concl}

The recent construction of an AME state of four subsystems
with six levels each  presented in  \cite{RBBRLZ22}
and further analyzed in \cite{ZBRBRL22}  
closes the last gap at $d=6$
and implies that AME$(4,d)$ states exist for any dimension $d\ge 3$.
As explained in Observation 1,  
the explicit construction of the state  AME$(4,6)$
implies that there exist: 

1) a solution QOLS$(6)$ to the quantum version of the problem of Euler,

2) a $2$-unitary matrix $U_{36}\in U(36)$,

3) a perfect tensor \cite{PYHP15} $ T_{ijk\ell}$ 
  with four indices, each running from $1$  to $6$,

4) a quantum orthogonal array  \cite{GRMZ18}
with $r=36$ rows,
$N_C=2$ classical columns and $N_Q=2$ quantum, 
and $d=6$-letters symbols of strength $k=2$,
written QOA$(36, 2_C+2_Q,6,2)$,

5) a quantum code  $((4, 1, 3))_6$
which allows one to encode one quhex in four such systems
 -- see \cite{GBR04}. 

Note that the existence of the golden $|{\rm AME}(4,6)\rangle$ state corresponding to
 the problem of  $36$  {\sl quantum officers} of Euler
 has some direct applications in processing of quantum information \cite{HCLRL12}.
 Consider a set of four dice, each with six faces, 
 prepared in such a maximally entangled state.
 Any pair of dice is unbiased, although 
 performing measurements on any selected pair of dice
 one can predict the outcome of the measurement 
 done on the other two. A similar protocol cannot work with four coins,
 as there are no AME states for four-qubit system.

The results mentioned above have also direct geometric implications.
Results presented in \cite{HS00} imply that  there are no absolutely 
maximally entangled states of a four qubit system $ABCD$.
Thus,  three copies of the projective group manifold $PU(4)$,
each consisting of pure states
maximally entangled with respect to three  different splittings,
 embedded in ${\mathbbm C}P^{15}$ 
{\sl do not}  intersect in a single point \cite{HS00}. 
However, for any dimension $d\ge 3$ such four-party entangled states do exist.
Therefore, if $d\ge 3$ then three copies of the manifold $PU(d^2)$, corresponding to three different  splittings of the four-qudit system $ABCD$, mutually intersect.
This non-empty set of intersection consists of $2$-unitary matrices from $U(d^2)$.

\begin{figure}[htbp]
\begin{center}
\includegraphics[width=0.48\textwidth]{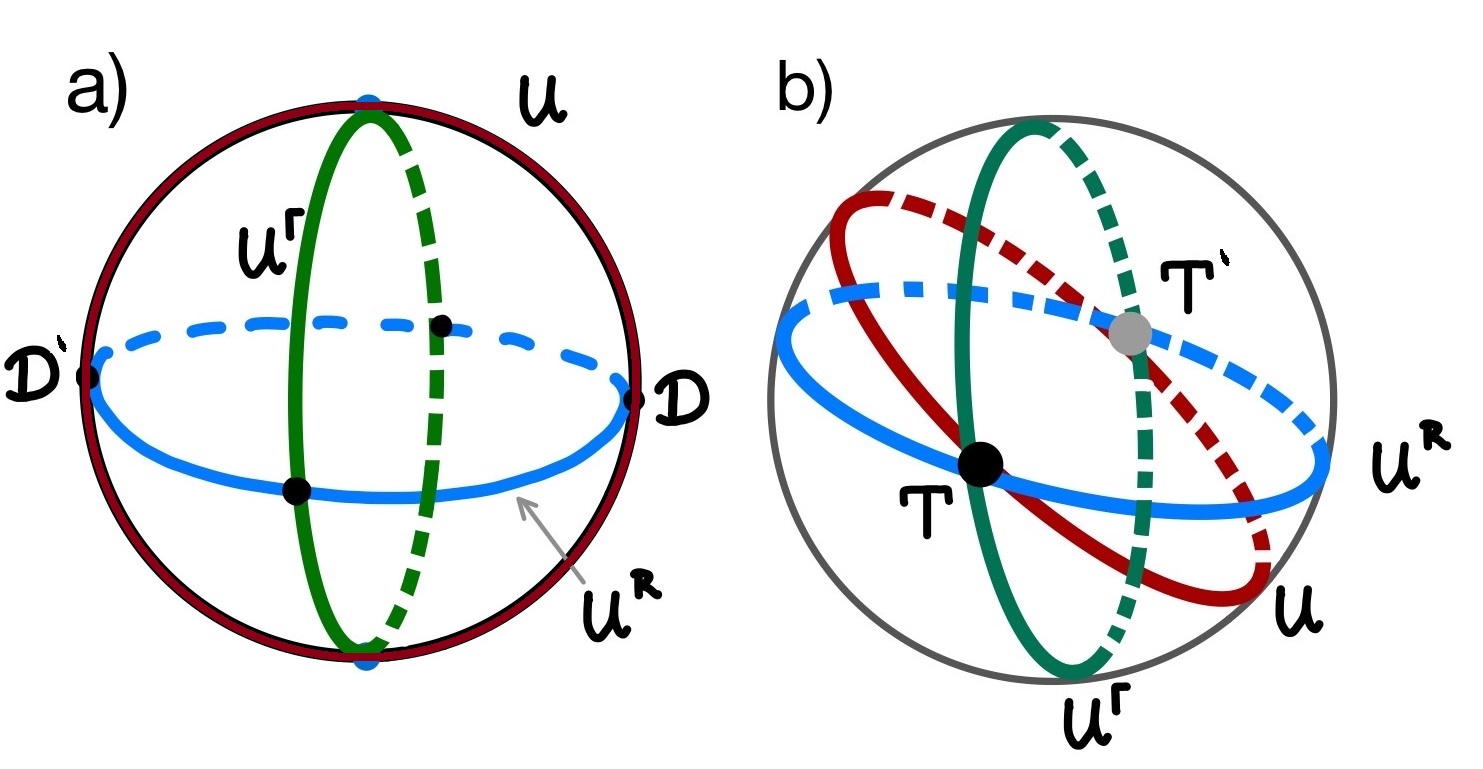} 
\caption{Intersection of manifolds 
${\cal P}_{\rm max}$  of maximally entangled states. 
a) $d=2$: manifold of maximally entangled states of two ququarts,
corresponding to the projective unitary group $PU(4)$,
embedded in ${\cal P}_{16} = {\mathbbm C}P^{15}$,
is represented by a great circle on the sphere.
Such three circles, each representing matrices such that $U$ or $U^R$ or $U^{\Gamma}$
are unitary, do not meet in a single point, as there are no $2$-unitary
matrices in $U(4)$.
 Since the manifold $\mathcal{P}_{\rm max}=U(4)/U(1)$ is
    conjectured to be non-displaceable, its two copies do intersect:
    point $D$ represents dual unitary matrices; b) $d \ge 3$: three
    copies if the manifold $\mathcal{P}_{\rm max}$, representing
     three different splittings of $ABCD$, do intersect at a point corresponding to a  two-unitary matrix $T$ in $U(d^2)$.}
\label{fig:three_circles}
\end{center}
\end{figure}

It is not simple to visualize mutual intersections of such highly dimensional manifolds.
Hence, in Fig. \ref{fig:three_circles} each projective group manifold  
$PU(d^2)$
 is represented  by a great circle $S^1$ on a sphere $S^2$.
In the case $d=2$ three such manifolds (great circles)
do not mutually intersect -- see  Fig. \ref{fig:three_circles}a.
Lack of the AME states for four-qubit systems can be related
to {\sl frustration} \cite{FFMPP10,FFMPP10b}: Requiring that certain 
two conditions hold, say matrices $U$ and $U^R$ are unitary,
implies that the third condition, $U^{\Gamma}\in U(4)$,
cannot be satisfied. In a geometric perspective, if a point belongs to
the intersection of two  circles, it cannot belong to the third one.
Fortunately frustration can be lifted if the numuber of degrees of freedom
increases. This is the case if one studies the same AME$(4,d)$ problem for any $d\ge 3$.
Then there exist a $2$-unitary matrix $U\in U(d^2)$ such that 
 $U^{\Gamma}$  and $U^{R}$ are unitary, 
 so that three circles meet in a single point 
 --  see  Fig. \ref{fig:three_circles}b).

It is easy to guess that such a simplistic visualization cannot
 display correctly all the features of multidimensional manifolds.
For instance, the set $D$ of dual unitary matrices  \cite{BKP19,BKP20}
formed by cross-section of two manifolds, corresponding to
$U$ and $U^R$ being unitary, does not consist of discrete points only, 
but forms a continuous set \cite{RAL2020,MPW22}.
As dual unitary matrices maximize the distance to the local product gates \cite{BRRL22}, 
in the space of bi-partite unitary gates they play the role of 
the maximally entangled states in the space of bi-partite pure quantum states.
Furthermore, the fact that three circles plotted in Fig. \ref{fig:three_circles}b)
intersect at the two-unitary matrix $T$
and also at the antipodal point $T'$ has no relation to the open problem
concerning the classification 
of AME states of a $N$ subsystems with $d$ levels each,   
which are not locally equivalent  \cite{BR20,RBB21}.

Let us emphasize that no real solution for the 
AME states of four subsystems with six levels each was found.
It was then conjectured \cite{RBBRLZ22}
that no real 2-unitary orthogonal matrices of size $36$
exist.  From a geometric perspective
it is thus interesting to analyze differences 
between the structures of various manifolds embedded in 
the complex and in the real projective spaces.
Frustration can also be relaxed if instead of considering 
standard quantum states one replaces complex coefficients,
by vectors with $n_c> 2$ real numbers \cite{FFPPS16}.

Note also that mutually orthogonal Latin cubes
 correspond to six-party states $|{\rm AME}(6,d)\rangle$ 
 and $3$-unitary matrices of size $d^3$,
 while $M$-dimensional Latin hypercubes 
 determine AME states of $N=2M$ subsystems
 and $M$-unitary matrices of order $d^M$,
 with the property  that all their reorderings
 remain unitary  \cite{GALR15}.
 It is clear that known combinatorial results
 concerning the existence of such orthogonal Latin cubes and hypercubes \cite{CD07},
 implies existence of certain multipartite entangled states
and can also be translated into the geometric
 language of intersection of the corresponding projective groups
 embedded in the complex projective space.

\medskip

It is a pleasure to thank  S. A. Rather, A.~Burchardt, W.~Bruzda, 
G.~Rajchel-Mieldzio{\'c} and A.~Lakshminarayan
for a long and fruitful collaboration on the quantum version of the
problem of Euler and to Z.~Pucha{\l}a and {\L}.~Rudnicki for mutual
research on mutually maximally entangled states.
The author is grateful to P.~Kielanowski for the invitation to a 
conference in Bia{\l}ystok, during which this work was presented,
and for an effective encouragement to complete this paper.
Furthermore, I am obliged to J.~Buczy{\'n}ski for organizing in Warsaw
a successful workshop {\sl Tensors from physics view point},
during which the final scope of this contribution was reshaped
in discussions with A.~Bor{\'o}wka and A.~Sawicki.
I am thankful to W. Bruzda for preparing Figs.
\ref{fig:OLS6} and  \ref{fig:QOLS6}, 
 to M. {\.Z}yczkowska for drawing the other figures
and to I. Bentgsson, M. Grassl, R.~Horodecki
  and S. Pascazio for numerous remarks on the manuscript.
This work was supported by Narodowe Centrum Nauki 
under the Quantera project number 2021/03/Y/ST2/00193 
and by Foundation for Polish Science 
under the Team-Net project no. POIR.04.04.00-00-17C1/18-00.

\appendix

\section{Glossary of key mathematical terms used}
\label{sec_glos}

\subsection{Algebraic}

Consider a matrix $A$ of order $d$ with entries $A_{ij}=\langle i|X|j\rangle$.
According to the standard notation $A^T$ 
represents the {\bf transposed} matrix with entries  $A_{ji}$,
while $A^{\dagger}$ denotes the Hermitian conjugate, 
$A^{\dagger}_{ij}={\bar {A_{ji}}}$.
A matrix $U$ is called {\sl unitary} if $UU^{\dagger}={\mathbbm I}_d$.

It is convenient to describe a bi-parite $d\times d$ state
in a composed Hilbert space ${\cal H}_d \otimes {\cal H}_d$.
Introducing a product basis $|a,i\rangle = |a\rangle \otimes |i\rangle$,
one can represent entries of a matrix $X$ of size $d^2$
in a $4$-index notation
$X_{\stackrel{\scriptstyle a i}{b j}} = \langle a,i |X|b,j\rangle$ -- see \cite{ZB04}.
Then the {\bf transposed} matrix reads,
$(X_{\stackrel{\scriptstyle a i}{b j}})^T =X_{\stackrel{\scriptstyle b j}{a i}}$.

\medskip

 {\bf A1.  Partial transpose} is obtained by action of the transpose on a single subsystem only.
Partially transposed matrix $X^{\Gamma}$ 
consist of entries transposed in each block,
$(X_{\stackrel{\scriptstyle a i}{b j}})^{\Gamma}=X_{\stackrel{\scriptstyle a j}{b i}}$,
so that the second pair of indices is exchanged.
There exist dual operation of partial transpose with respect to the first subsystem,
$(X_{\stackrel{\scriptstyle a i}{b j}})^{\urcorner}=X_{\stackrel{\scriptstyle b i }{a j}}$,
and their concatenation produces the standard transpose,
$(X^{\Gamma})^{\urcorner}= (X^{\urcorner})^{\Gamma}=X^T$.

\smallskip

{\bf A2.  Reshuffling} $X^R$ of a matrix  $X$ of size $d^2$, (also called {\sl reordering}),
consists in reshaping rows of a matrix of length $d^2$ into square blocks of size $d$.
This corresponds to a diagonal swap of the indices,
$(X_{\stackrel{\scriptstyle a i}{b j}})^{R}=X_{\stackrel{\scriptstyle a b}{i j}}$,
and the dual transformation producing $X_{\stackrel{\scriptstyle j i}{b a}}$
has similar properties \cite{ZB04}.

\smallskip

{\bf A3.  Dual unitary} matrix $X$ of size $d^2$ satisfies {\sl two} conditions:
     a) it is unitary, $X\in U(d^2)$, and b) its reshuffling is also unitary, 
         $X^R\in U(d^2)$ -- see \cite{BKP19,ARL21}.
        Such matrices are represented in Fig. \ref{fig:three_circles}a
         by  the intersection of two circles.

\smallskip

{\bf A4.  Two-unitary} matrix $X$ of size $d^2$,
            introduced earlier in \cite{GALR15}, satisfies {\sl three} conditions:
              a) $X\in U(d^2)$, 
              b) $X^R\in U(d^2)$, 
              c) $X^{\Gamma}\in U(d^2)$. 
            Thus this property is stronger as the previous one
                 and any $2$-unitary $U$ is also dual-unitary.
             Two-unitary matrices correspond in Fig. \ref{fig:three_circles}b
                to the intersection of three circles.
  An example of a $2$-unitary permutation matrix $P_9$ of order nine
  is obtained by replacing each card in Eq. (\ref{P9}) by digit $1$.

\smallskip

{\bf A5.  Multi-unitary} matrix $X$ of size $d^M$, labeled by $2M$ indices,
                is unitary, $X\in U(d^M)$, 
                and it remains unitary under reordering of its entries
                corresponding to an arbitrary symmetric partition of $2M$ indices  \cite{GALR15}.
                The number of such partitions is ${2M\choose M}$.
                 Since we know that if $U$ is unitary so is $U^T$,
                  the number of conditions to be checked to establish 
                $M$-unitarity is thus  $\frac{1}{2}{2M\choose M}$, 
                 which gives $3$ conditions for $2$-unitarity and
                 $10$ conditions for $3$-unitarity.
             
\smallskip

{\bf A6.  Flattening}  of a tensor with more than two indices
   denotes its reshaping into a matrix.    
 A  tensor $T_{ijk\ell}$, with four indices, each running from $1$ do $d$,
  can be transformed into a matrix in three ways, corresponding to
 projections of this hypercube into three different planes.
 Defining two composite indices,
 $\mu=(i-1)d+j$ and $\nu=(k-1)d+\ell$,
 we can represent the tensor as a matrix od size $d^2$
 with entries $X_{\mu\nu}=T_{ij,k\ell}$.
 If the index $\mu$ is composed by the pair  $(i,k)$
 one obtains another flattening, $Y_{\mu\nu}=T_{ik,j\ell}$,
 while the third choice,  $(i,\ell)$, leads to  the matrix $Z_{\mu\nu}=T_{i\ell,kj}$.
 All three flattenings are related by the reorderings defined above,
 as $Y=X^R$ and $Z=X^{\Gamma}$.

\smallskip

{\bf A7.  Perfect tensor} has the property that all its flattenings are unitary up to a constant \cite{PYHP15}.
Hence a four-index tensor $T_{ijk\ell}$ restructured into a matrix
gives a $2$-unitary matrix 
e.g. $X_{\mu\nu}=T_{ij,k\ell}$, such that all its reorderings remain unitary.

\medskip
\subsection{Combinatorial}

{\bf B1.  Latin square} of order $d$, written as  LS(d),
   consists of $d$ copies of $d$ symbols,
 arranged into a square such that each row and each 
 column contains different symbols. These configurations
 exist for any  dimension $d$ \cite{CD07}.

\smallskip
 
{\bf B2.  Graeco-Latin square} of order $d$, 
also called a pair of {\sl orthogonal Latin squares},  written OLS(d),
are formed by two Latin squares which are {\sl orthogonal}.
This means that all $d^2$ pairs of symbols placed in each cell of the square
are different -- see conditions C1)-- C3) in Sec. \ref{sec_quantOLS}.
 These configurations exist \cite{CD07} for $d=3,4,5$ and $d\ge 7$.
 A collection of $m$ LS of order $d$ 
 is called mutually orthogonal Latin squares, denoted as $m$-MOLS(d),
 if any two of them are orthogonal. There exist not more than $(d-1)$ MOLS$(d)$
 and this bound is saturated for dimension $d$ being a prime or a power of prime \cite{JCD01}.
  
\smallskip 
 
 {\bf B3.  Latin cube} of order $d$, written as  LC(d),
   consists of $d^2$ copies of $d$ symbols,
 arranged into a cube such that each row,
  each  column and each line contain each of the symbols exactly once.
 In analogy to OLS one defines also orthogonal Latin cubes 
and orthogonal Latin hypercubes of an arbitrary dimension $n$ \cite{MKW08}.
Latin cubes are used in coding theory.

\smallskip

{\bf B4.  Orthogonal array} with $r$ runs, $N$ factors,
  $d$ levels and strength $k$, written OA$(r,N,d,k)$,
  forms a rectangular $r \times N$ table with entries 
  coming from a $d$-letter alphabet arranged in such a way
  that in every collection of $k$ columns each of the possible ordered pairs of 
  elements appears the same number of times \cite{St03}. 
  Observe that  LS$(d)$ is equivalent to OA$(d^2,3,d,2)$,
  while  LC$(d)$ to OA$(d^3,4,d,3)$.
   In general, OA$(d^2, 2+m,d,2)$ defines a set of
    $m$-MOLS$(d)$.
  
\smallskip

{\bf B5.  Quantum Latin square} of order $d$, written QLS$(d)$,
is a collection of $d^2$ quantum states, $|\psi_{ij}\rangle \in {\cal H}_d$,
 arranged into a square such that in each row and each column
 all states are orthogonal and form an orthogonal basis \cite{MV16}.
 If the number of different states in the design exceeds $d$
 then the design is genuinely quantum \cite{PWRBZ21}, 
 as it cannot be transformed  into a classical LS by a unitary rotation.
 
\smallskip
 
 {\bf B6.  Quantum orthogonal Latin squares} of order $d$, 
 written  QOLS$(d)$,
 is a collection of $d^2$ bi-partite quantum states 
 $|\psi_{ij}\rangle \in
  {\cal H}_d \otimes  {\cal H}_d$
 arranged into a square of size $d$  such that 
conditions Q1)--Q3) listed in Sec. \ref{sec_quantOLS}
are satisfied \cite{Ri20}.
Such designs exists for any $d\ge 3$ \cite{RBBRLZ22}.

\smallskip

{\bf B7.  Quantum orthogonal array}  QOA$(r,N=N_C+N_Q,d,k)$
is an arrangement consisting of $r$ rows composed of
$N$-partite quantum states with $d$ levels each,
$|\phi_j\rangle \in {\cal H}_d^{\otimes N}$,
such that for any subset ${\cal I}$ consisting of $(N-k)$
subsystems the following partial trace relation holds \cite{GRMZ18},
$\sum_{j=1}^N {\rm Tr}_{\cal I} 
(|\phi_j\rangle \langle \phi_j|) \propto \mathbbm{I}_{d^k}$.
Out of $N$ columns of the entire design one distinguishes a certain number $N_C$ 
of classical columns containing product states and the remaining $N_Q=N-N_C$ quantum
columns consisting of entangled states. Classical OA correspond to the case
$N_C=N$ and $N_Q=0$.
The notion of QOA forms a natural generalization
of other quantum designs.
For instance, QOA$(8,3_C+3_Q,2,3)$ is equivalent to 
3QOLC$(2)$, which corresponds to  the state AME$(6,2)\in {\cal H}_2^{\otimes 6}$ \cite{GRMZ18},
while QOA$(36,2_C+2_Q,6,2)$ gives 2QOLS$(6)$
and the state AME$(4,6)\in {\cal H}_6^{\otimes 4}$ \cite{RBBRLZ22}.

\subsection{Geometric}

{\bf G1.  Manifold}  $\cal M$  is a topological space locally equivalent 
to Euclidean space. Examples include circle, sphere, and projective space. 
These examples form closed, compact and connected 
 $n$-manifolds of a fixed dimension $n$.

\smallskip

{\bf G2.  Real projective space}  ${\mathbbm R}P^n$ --
  a compact, smooth manifold of dimension $n$
 defined as a space of lines passing through the origin in ${\mathbbm R}^{n+1}$.
 Examples include
  ${\mathbbm R}P^1$ topologically equivalent to the circle $S^1$,
   ${\mathbbm R}P^2$, called real projective plane,
    which can be obtained from the sphere $S^2$ by
      identifying antipodal points and
   ${\mathbbm R}P^3$, diffeomorphic to $SO(3)$ and $U(2)/U(1)$.

\smallskip

{\bf G3.  Complex projective space}  ${\mathbbm C}P^n$ 
is the projective space with respect to the field of complex numbers.
It forms a complex, compact, smooth manifold of $2n$ real dimension
and can be regarded as a quotient space,
${\mathbbm C}P^n=S^{2n+1}/U(1)$ 
so that ${\mathbbm C}P^1=S^{3}/S^1=S^2.$ 
  
\smallskip
 
{\bf G4.  Symplectic manifold}  is a smooth manifold $\cal M$, 
equipped with a  closed nondegenerate differential $2$-form $\omega$, 
called the symplectic form. Symplectic manifold is used in 
classical mechanics to represent the phase space.

\smallskip

{\bf G5. Lagrangian manifold} ${\cal L}_n$ of dimension $n$
is a differentiable submanifold of a $2n$-dimensional symplectic manifold ${\cal M}_{2n}$
such that the exterior form $\omega$ specifying the symplectic structure
 vanishes identically on  ${\cal L}_n$.

\smallskip
 
{\bf G6.  Displaced manifold}. 
An image by a transformation determined by a unitary matrix $U$ of a suitable dimension $m$
 of an $n$-dimensional manifold ${\cal M}$ embedded inside $\cal N$. 
A displaced manifold will be written $U{\cal M} \subset {\cal N}$.

\smallskip

{\bf G7.  Non-displaceable manifold} ${\cal M}$
embedded  in ${\cal N}$
such that ${\cal M}$ cannot be displaced unitarily in such a way that 
it does not intersect its image: 
For any $U\in U(m)$  one has
 ${\cal M} \cap U{\cal M} \ne {\emptyset}$.

\smallskip

{\bf G8. Foliation} of an $n$ dimensional manifold ${\cal M}_n$ is
its decomposition into a union of disjoint {\sl leaves} --
connected submanifolds of dimension $p$, which can be generated by equivalence classes.
The number $q=n-p$, called codimension, detemines the dimension
of the space of leaves. A foliation is called  {\sl singular}
if the dimension of the leaves is allowed not to be constant.
Singular foliation of ${\mathbbm C}P^{3}$ visualized in Fig. \ref{fig:2qubit}
consists of generic leaves  of $5$ real dimensions, so the codimension $q$
reads $6-5=1$.
In foliation of  the $16$-dimensional manifold ${\mathbbm C}P^{8}$,
presented in  Fig. \ref{fig:2qutrit},
the generic orbit has $p=14$ real dimensions
and its codimension,  $q=16-14=2$, 
is equal to the dimension of the triangle of the Schmidt vectors.

\end{document}